\documentclass[journal,twoside,web]{ieeecolor}
\usepackage{amsmath,amssymb,amsfonts}
\usepackage{algorithm}
\usepackage{algorithmicx}
\usepackage{algpseudocode}
\usepackage{array}
\usepackage{bm}
\usepackage{booktabs}
\usepackage{cite}
\usepackage{graphicx}
\usepackage{tmi}
\usepackage{tabularx}
\usepackage{arydshln}
\usepackage{subcaption}
\usepackage{textcomp}
\usepackage{url}
\newcommand{\circnum}[1]{\textcircled{\small #1}}
\newcolumntype{L}{>{\raggedright\arraybackslash}X}
\newcolumntype{Y}{>{\centering\arraybackslash}X}
\newcolumntype{C}{>{\centering\arraybackslash}X}
\def\BibTeX{{\rm B\kern-.05em{\sc i\kern-.025em b}\kern-.08em
    T\kern-.1667em\lower.7ex\hbox{E}\kern-.125emX}}
\begin{document}
\title{MorphiNet: A Graph Subdivision Network for Adaptive Bi-ventricle Surface Reconstruction}
\author{Yu Deng, Yiyang Xu, Linglong Qian, Charlène Mauger, Anastasia Nasopoulou, Steven Williams, Michelle Williams, Steven Niederer, David Newby, Andrew McCulloch, Jeff Omens, Kuberan Pushprajah, Alistair Young
\thanks{This paper was first submitted for review on 31st Dec 2024. 
YD is funded by the Kings-China Scholarship Council PhD Scholarship Program. YX is funded by the EPSRC Centre for Doctoral Training in Smart Medical Imaging (EP/S022104/1) and the Wellcome/EPSRC Centre for Medical Engineering (WT203148/Z/16/Z). YX acknowledges stipend funding contribution from Heartflow Inc., CA, USA. AM, JO, AY, AN, CM, and KP acknowledge funding from the National Institutes of Health (R01HL121754). SW acknowledge the funding from the British Heart Foundation (FS/20/26/34952). Infrastructure was supported by the Wellcome/EPSRC Centre for Medical Engineering (WT203148/Z/16/Z). SCOT-HEART was funded by the Chief Scientist Office of the Scottish Government Health and Social Care Directorates (CZH/4/588), with supplementary awards from Edinburgh and Lothian’s Health Foundation Trust and the Heart Diseases Research Fund.}
\thanks{Yu Deng, Yiyang Xu, Charlène Mauger, Anastasia Nasopoulou, Kuberan Pushprajah, Alistair Young are with the School of Biomedical Engineering and Imaging Sciences, King's College London, UK (e-mail: yu.deng@kcl.ac.uk; yiyang.1.xu@kcl.ac.uk; charlene.1.mauger@kcl.ac.uk; anastasia.nasopoulou@kcl.ac.uk; kuberan.pushparajah@kcl.ac.uk; alistair.young@kcl.ac.uk).}
\thanks{Yu Deng is also with the Brain Imaging \& Neuro Epidemiology Department, Luxemboug Institute of Health (e-mail: yu.deng@lih.lu).}
\thanks{Linglong Qian is with the Department of Biostatistics and Health Informatics, King's College London (e-mail: linglong.qian@kcl.ac.uk).}
\thanks{Steven Williams, Michelle Williams and David Newby are with the Centre for Cardiovascular Science, University of Edinburgh, UK (e-mail: steven.williams@ed.ac.uk; michelle.williams@ed.ac.uk; d.e.newby@ed.ac.uk).}
\thanks{Andrew McCulloch and Jeff Omens are with the Department of Bioengineering, University of California, San Diego (email: amcculloch@ucsd.edu; jomens@ucsd.edu).}
\thanks{Steven Niederer is with the National Heart and Lung Institute (NHLI), Imperial College London, UK (e-mail: s.niederer@imperial.ac.uk).}
}
\maketitle

\begin{abstract}
Cardiac Magnetic Resonance (CMR) imaging is widely used for heart model reconstruction and digital twin computational analysis because of its ability to visualize soft tissues and capture dynamic functions. However, CMR images have an anisotropic nature, characterized by large inter-slice distances and misalignments from cardiac motion. These limitations result in data loss and measurement inaccuracies, hindering the capture of detailed anatomical structures. In this work, we introduce MorphiNet, a novel network that reproduces heart anatomy learned from high-resolution Computed Tomography (CT) images, unpaired with CMR images. MorphiNet encodes the anatomical structure as gradient fields, deforming template meshes into patient-specific geometries. A multilayer graph subdivision network refines these geometries while maintaining a dense point correspondence, suitable for computational analysis. MorphiNet achieved state-of-the-art bi-ventricular myocardium reconstruction on CMR patients with tetralogy of Fallot with 0.3 higher Dice score and 2.6 lower Hausdorff distance compared to the best existing template-based methods. While matching the anatomical fidelity of comparable neural implicit function methods, MorphiNet delivered 50$\times$ faster inference. Cross-dataset validation on the Automated Cardiac Diagnosis Challenge confirmed robust generalization, achieving a 0.7 Dice score with 30\% improvement over previous template-based approaches. We validate our anatomical learning approach through the successful restoration of missing cardiac structures and demonstrate significant improvement over standard Loop subdivision. Motion tracking experiments further confirm MorphiNet's capability for cardiac function analysis, including accurate ejection fraction calculation that correctly identifies myocardial dysfunction in tetralogy of Fallot patients. Code and checkpoints are available at \url{https://github.com/MalikTeng/MorphiNetV2}.
\end{abstract}


\section{Introduction}
\label{sec:introduction}
\IEEEPARstart{H}{eart} anatomy is commonly represented by 3D mesh models reconstructed from medical imaging. Mesh models are valuable for investigating cardiac biomechanics and electrophysiology and for evaluating treatments through simulations that replicate heart characteristics \cite{chabiniok2016multiphysics,Nasopoulou2017}. Cardiovascular Magnetic Resonance (CMR) imaging is useful for heart model reconstruction because it can visualize soft tissues, provide detailed anatomical structural information, and capture dynamic cardiac functions with high temporal resolution and signal-to-noise ratio without ionizing radiation. However, accurate model reconstruction from CMR data is challenging because of its anisotropic nature. This anisotropy presents as large inter-slice distances which result from the trade-off between relatively slow data acquisition and scanning time constraints. Moreover, differences in breath hold positioning between slices (that is, motion artifacts) can further misalign the data, ultimately producing incomplete and inaccurately registered information that does not convey the full anatomy of the heart \cite{villard2017correction}. Although fully volumetric (3D) CMR acquisitions can mitigate slice misalignment issues, it is not an imaging format that is routinely used in clinical practice, and these challenges persist. 

The critical step in heart model reconstruction is to obtain precise anatomical structures and dimensions. For example, inaccuracies in the dimensions of the cavity or the myocardial wall can result in misdiagnosis of conditions such as heart failure or hypertrophy. Although super-resolution and advanced interpolation techniques offer improvements \cite{tarroni2018comprehensive,xia2021super}, these methods often struggle to capture the full variability of heart anatomy and motion and ultimately affect the accuracy of functional analyses \cite{gong2022robust}.

Mesh generation approaches can be broadly classified into \textbf{template-based methods} and \textbf{neural implicit function methods}. Template-based methods define surfaces using vertices, edges, and faces to form a structured template mesh with dense point correspondence, providing an efficient representation of the physiological characteristics of the heart compatible with finite element analysis \cite{hughes2012finite}. These methods are computationally efficient, allowing single-pass inference. One early approach learned neighborhood sampling and mesh unpooling in a neural network structure that deformed a template mesh directly to volumetric images \cite{wickramasinghe2020voxel2mesh}. Similarly, using graph convolution networks to deform whole-heart mesh templates via biharmonic coordinates makes possible the creation of simulation-ready heart models in an end-to-end manner \cite{kong2022learn}. 

A similar multi-modal approach combined voxel processing with super-resolution in a graph convolution network that successfully reconstructed 3D+t heart models from CMR images \cite{deng2023modusgraph}. In contrast, neural implicit function methods represent surfaces through level sets or signed distance fields \cite{osher2004level}, requiring iterative optimization processes that result in substantially longer computational inference times, but can offer high geometric fidelity through pixel-wise optimization. One approach is to learn coordinate-to-label set mappings to perform continuous domain CMR segmentation \cite{stolt2023nisf}, and apply post-processing to create heart models. Another method used a signed distance field template combined with diffeomorphic flows to create heart models without post-processing \cite{sun2022topology}. A more direct approach enabled reconstruction of the heart model from sparse point clouds, using the DeepSDF framework \cite{park_2019_cvpr} trained solely on signed distance fields \cite{verhulsdonk2023shape}. 

However, these methods still suffer if there are missing cardiac structures, particularly at the base, and the distortion between slices leads to suboptimal geometric fidelity and variability of precision between datasets. Neural implicit functions offer high fidelity but pose a large amount of computation overhead and require dedicated post-processing to make generated meshes simulation ready, and are currently prohibitive for clinical applications. On the other hand, template-based approaches are inadequate because their deformation processes do not guarantee accurate anatomical structures. These processes are optimized by minimizing the Chamfer distance \cite{fan2017point} between a deformed mesh and the ground truth mesh or the point cloud. Although this reduces the average distance from each mesh vertex to its nearest ground truth point, it does not ensure accurate curvature or surface continuity given the data misses or distortions in the ground truth. Moreover, ground-truth 3D meshes are unavailable from CMR segmentation due to sparse through-plane resolution. 

\begin{figure}[!t]
    \centering
    \includegraphics[width=\columnwidth]{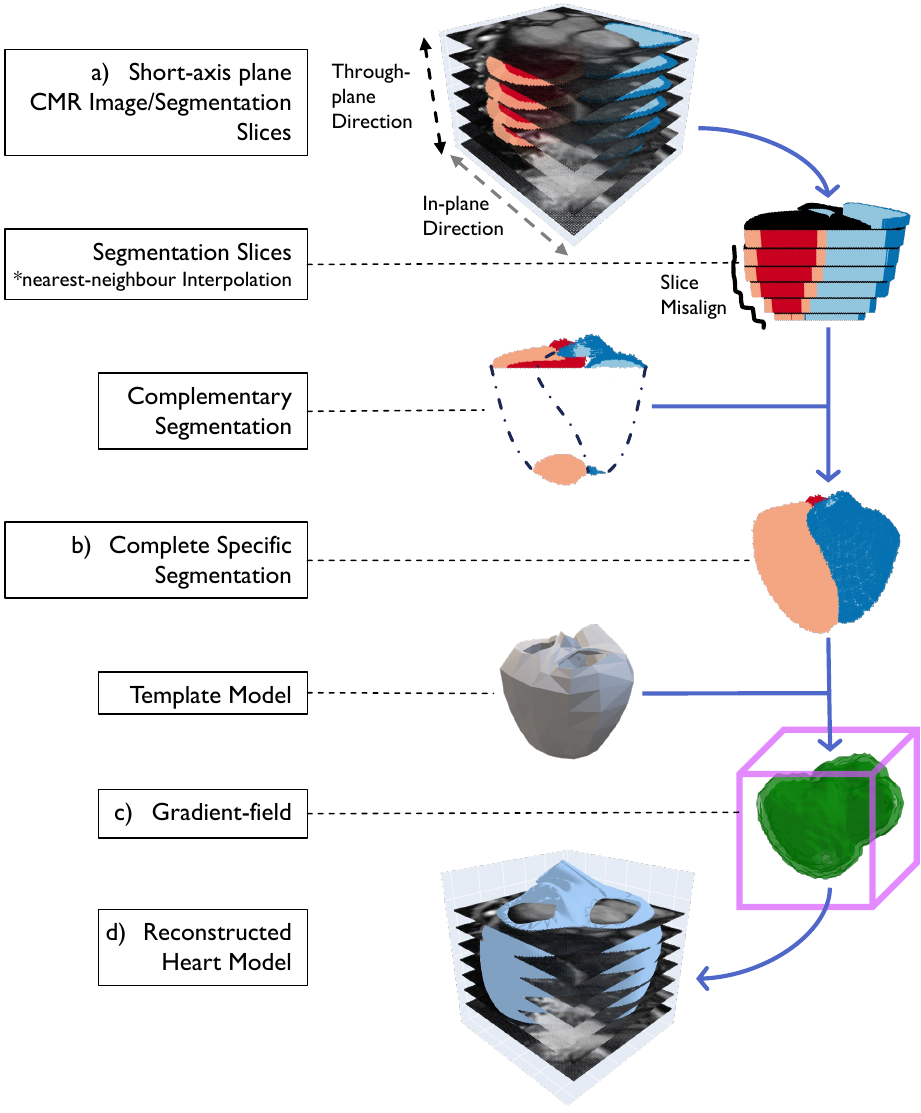}
    \caption{Overview of the proposed MorphiNet framework. From top to down: a) Segment the cardiac magnetic resonance images with large inter-slice distances and misalignments. b) Complete specific segmentation. c) Find a gradient field from the segmentation to adjust the template mesh. d) End with a reconstructed heart model preserving underlying cardiac anatomy while conforming to patient-specific data.}
    \label{fig:graphical_summary}
\end{figure}

In this work, we address these challenges by proposing \textbf{MorphiNet}, an efficient method to restore missing structural information about the heart, and create high quality morphologies, as shown in Fig \ref{fig:graphical_summary}. MorphiNet learns the true underlying anatomy as a \textit{complementary segmentation} and uses it to \textit{complete a specific segmentation}. We realized that such anatomy naturally conveyed from Computed Tomography (CT) imaging because of its higher spatial resolution. Given unpaired CMR and CT image data, MorphiNet learns the complementary segmentation, transfers it to CMR data, and automatically performs patient-specific tuning to generate complete segmentation. MorphiNet decodes the specific segmentation as a gradient field, such that deforming the template mesh conforms to heart geometries. It then uses adaptive subdivision with a Graph Subdivision Network (GSN) to refine geometries in complex surface areas while maintaining a dense point correspondence. MorphiNet offers an interpretable and efficient solution for incorporating accurate anatomical structures into heart model reconstructions. Unlike previous methods that function as black-box systems, MorphiNet provides transparent vertex updating and face subdivision mechanisms through well-defined stages, enhancing reliability assessment in clinical applications.

We focus on the reconstruction of the Left Ventricular (LV) and Right Ventricular (RV) myocardium, which represents the most challenging aspect of cardiac reconstruction and is particularly valuable for biomechanical and electrophysiological simulation applications. Evaluating our distinct network architecture against current state-of-the-art template-based methods, MorphiNet obtained the best anatomical fidelity. Compared with neural implicit function methods, MorphiNet shows similar anatomical fidelity with substantially better computational efficiency. 

We further validated the effectiveness of MorphiNet to learn the anatomical structure by first masking parts of the heart anatomy in the input segmentation and evaluating the completeness and structural correctness of the predicted missing parts. We also demonstrated the effectiveness of gradient field-based deformation and learned graph subdivision by MorphiNet compared to the default Loop surface subdivision algorithm \cite{loop1987smooth}. Moreover, we show that the MorphiNet generation is capable of tracking motion with a faithful representation of cardiac motion and estimation of ejection fraction that correctly indicates myocardial dysfunction in patients with tetralogy of Fallot. Our main contributions are as follows.
\begin{itemize}
    \item We introduce MorphiNet, the first approach to address anisotropic CMR challenges by learning cardiac anatomy from unpaired CT data, enabling bi-ventricular myocardium reconstruction without requiring CMR ground-truth 3D meshes for training.
    \item We propose a novel gradient field representation that transforms template meshes into patient-specific geometries through an interpretable deformation process, overcoming limitations of previous approaches.
    \item We develop a GSN that adaptively refines mesh geometries while maintaining dense point correspondence, crucial for cardiac motion tracking and biomechanical analysis applications.
    \item We demonstrate superior anatomical fidelity across four datasets, achieving approximately 30\% higher Dice scores compared to the state-of-the-art template-based method, with clinical feasibility through efficient inference compared to neural implicit function methods.
    \item We validate cardiac assessment capabilities through motion-tracking experiments, demonstrating preserved mesh topology essential for dynamic cardiac analysis.
\end{itemize}

This paper extends our previous work \cite{deng2024adaptive} in the following ways: we introduce updates to the methodology incorporating dynUNet for segmentation and ResNet for completing specific segmentation, providing a detailed and transparent pipeline description. We elaborate more on adaptive mesh refinement using a GSN. We describe its multi-layer perceptron-based vertex adjustment and present mathematical formulations for template deformation and subdivision. The experimental section is expanded with an additional dataset, the Automatic Cardiac Diagnosis Challenge (ACDC), and we update experimental results and evaluation scores to fully assess the quality of the mesh. Detailed visualizations, including surface error maps and cross-sectional overlays, enhance the presentation of results. At the same time, extensive ablation studies and a more granular comparative analysis against state-of-the-art methods highlight MorphiNet's performance and contributions.

\section{Related Work}
\subsection{Template-Based Methods} 
Template-based methods define surfaces using vertices, edges, and faces to form a template mesh, which is deformed to the patient-specific images or point clouds, thereby providing structured representations with dense point correspondence. Young \emph{et al.} used interactive 3D finite element model customization to streamline the calculation of left ventricular mass and volume \cite{young2000left}. More automated approaches extended to multiple chambers and the development of 3D statistical shape models (SSMs) \cite{frangi2002automatic}. The introduction of 4D atlases by Perperidis \emph{et al.} captured dynamic cardiac changes and distinguished between inter- and intra-subject variability \cite{perperidis2005construction}. SSMs are essential for computational cardiac atlases, enabling personalized medicine and epidemiological studies of biophysical properties \cite{young2009computational}. Large-scale studies by Medrano-Gracia \emph{et al.} \cite{medrano2013large}, Bai \emph{et al.} \cite{bai2015bi}, and Mauger \emph{et al.} \cite{mauger2019right} demonstrated the value of extensive datasets, constructing detailed atlases that enable analysis of population shape variations. Integration of multiple imaging modalities, by Puyol-Antón \emph{et al.} \cite{puyol2017multimodal}, and the application of SSM for myocardial infarction classification, by Suinesiaputra \emph{et al.} \cite{suinesiaputra2017statistical}, improved clinical applicability. Gilbert \emph{et al.} reviewed machine learning in cardiac atlasing to improve automation and precision \cite{gilbert2020artificial}. Recent advances include applications in pediatric cardiology, with Marciniak \emph{et al.} \cite{marciniak2022three} identifying morphological changes linked to childhood obesity and Govil \emph{et al.} \cite{govil2023deep} achieving fully automated modeling for congenital heart disease. Although traditional atlas-based methods can achieve high accuracy and mesh quality, they do not scale well to large cohorts due to substantial computation time needed for automatic patient-specific customization. 

Deep learning-based template-based methods have emerged as powerful alternatives to traditional statistical approaches. Xu \emph{et al.} reformulated the problem through volumetric mapping, enabling flexible handling of multi-orientation contours without mesh constraints \cite{xu2019ventricle}. Wickramasinghe \emph{et al.} proposed Voxel2Mesh, an end-to-end framework that directly maps volumes to meshes using learned sampling and adaptive unpooling, yielding higher IoU with fewer vertices  \cite{wickramasinghe2020voxel2mesh}. Kong and Shadden proposed HeartDeformNet to predict control handle displacements to deform whole heart mesh templates via biharmonic coordinates, enabling CFD simulations of cardiac flow with temporally consistent 4D meshes \cite{kong2022learn}. Addressing specific clinical needs, Kong \emph{et al.} developed shape-disentangled modeling for congenital heart defects with virtual cohort generation capabilities \cite{kong2024sdf4chd}. Meng \emph{et al.} \cite{meng2023deepmesh} developed region-specific mesh tracking for cardiac dynamics, while Deng \emph{et al.} \cite{deng2023modusgraph} introduced ModusGraph, a voxel-guided framework, for accurate 3D/4D heart mesh reconstruction from sparse cine CMR. Taking a different approach, Chen \emph{et al.} focused on handling sparse point clouds using learned deformation registration \cite{chen2021shape}. Beetz \emph{et al.} introduced point cloud completion networks, achieving significant error reduction and cross-domain adaptability in public datasets \cite{beetz2023multi}. Although these methods can achieve dense point correspondence for computational analyses, they are sensitive to missing structures in the image or point cloud inputs. 

\subsection{Neural Implicit Functions} 
Neural implicit functions are used to represent surfaces through neural fields, level sets, or signed distance functions, offering flexibility in handling topological variations but requiring iterative optimization during inference. Mesh models can then be generated from marching cubes, providing high fidelity. Sun \emph{et al.} introduced Neural Diffeomorphic Flow (NDF), using neural ODE–driven diffeomorphic flows to build topology-preserving implicit templates for 3D organ reconstruction and registration across diverse datasets \cite{sun2022topology}. Stolt-Ansó \emph{et al.} presented NISF which learned continuous coordinate-to-label mappings with subject-specific priors, showing strong generalization and interpolation in UK Biobank 3D+t MRI data \cite{stolt2023nisf}. Verhülsdonk \emph{et al.} proposed a Lipschitz-regularized DeepSDF approach for bi-ventricular reconstruction from sparse point clouds, overcoming limitations of SSMs and segmentation-based pipelines \cite{verhulsdonk2023shape}. Muffoletto \emph{et al.} demonstrated robust reconstruction from limited standard CMR views \cite{muffoletto2023neural}. Most recently, Kong \emph{et al.} introduced a shape-disentangled signed distance function framework for congenital heart defects with virtual cohort generation capabilities \cite{kong2024sdf4chd}. Although anatomically accurate, inference optimization can be computationally intensive, and many methods do not enable dense point correspondence needed for SSM and biomechanical analysis.

\section{Method}
\begin{figure*}[!t]
    \centering{\includegraphics[width=\textwidth]{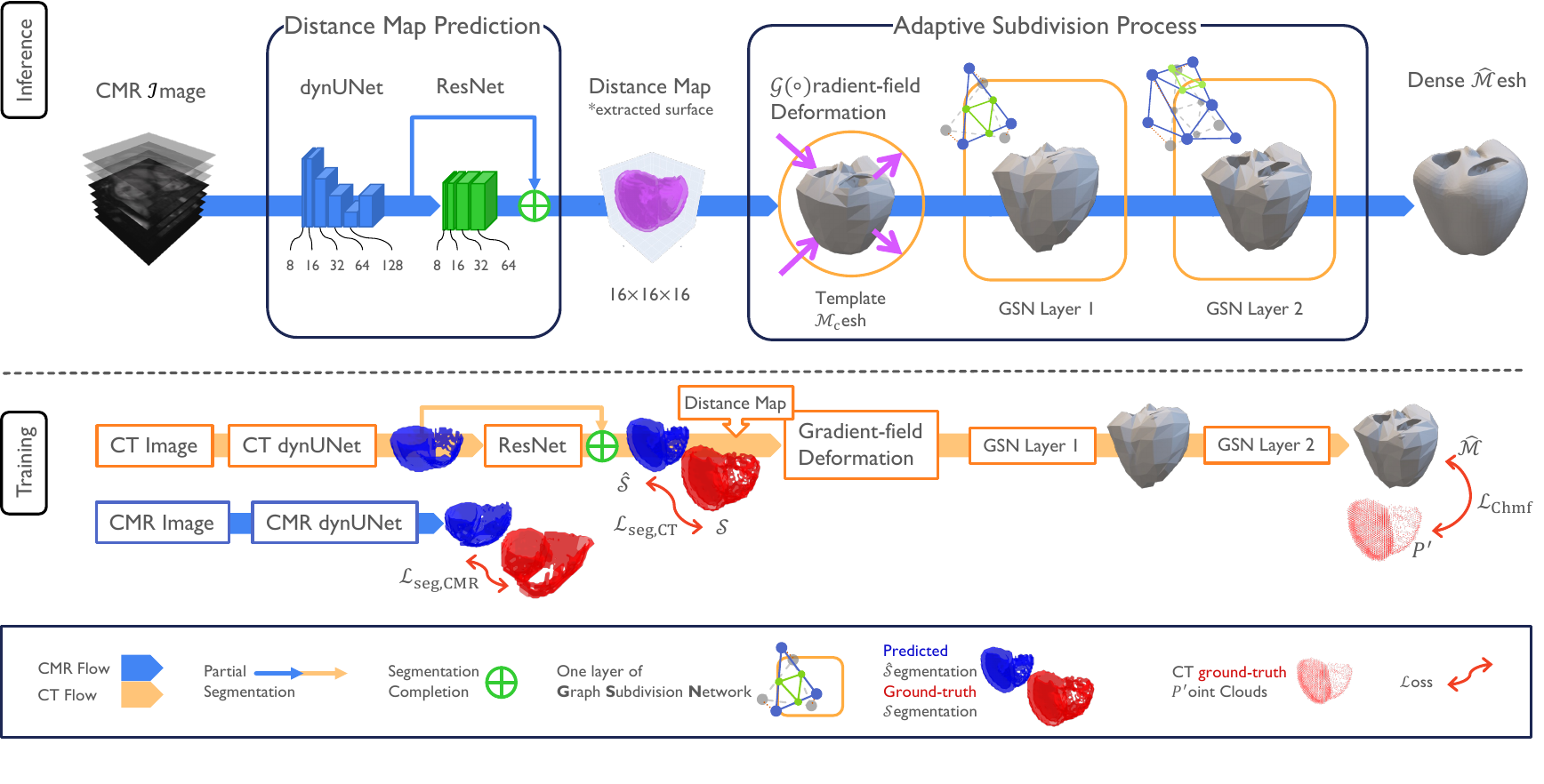}}
    \caption{\textbf{Diagram showing MorphiNet's training and inference workflows.} MorphiNet consists of three neural network modules with learnable parameters: dynUNet, ResNet, and GSN. \textit{Training phase}: CT images follow CT dynUNet$\ \rightarrow\ $ResNet$\ \rightarrow\ $GSN pathway; CMR images follow CMR dynUNet pathway only. \textit{Inference phase}: CMR images are processed through CMR dynUNet$\ \rightarrow\ $ResNet$\ \rightarrow\ $GSN when model parameters are frozen. CT inference follows CT dynUNet$\ \rightarrow\ $ResNet$\ \rightarrow\ $GSN pathway.}
    \label{fig:overview}
\end{figure*}

\textbf{MorphiNet} is a fully automatic, end-to-end pipeline to generate customized 3D mesh models, shown as the blue workflow in Fig \ref{fig:overview}. Taking a stack of short-axis (SAX) CMR images as input, a dynUNet segmentation network \cite{ranzini2021monaifbs} (a MONAI \cite{monai2020monai} implementation of nnU-Net \cite{isensee2021nnu}) infers the bi-ventricular myocardium region from the images. A ResNet decoder \cite{myronenko20193d} predicts the complementary segmentation, which is the segmentation near the basal and apex plane that is missing from the CMR stack. The completed segmentation is transformed into a distance map. A gradient field is derived from the distance map and is used to deform a template mesh, adjusting the mesh to conform to the actual anatomical structure in the patient data input. Following this, an \textbf{Adaptive Subdivision Process} with GSN layers refines the adjusted template mesh by increasing the number of surface points and encouraging finer surface adjustments. The result is a 3D mesh model with a smooth surface and densely corresponding points.

\subsection{Distance Map Prediction}
\label{sec:distance_map}
We denote cardiac images as a signal intensity function $\mathcal{I}:\mathbf{\Omega}\mapsto\mathbb{R}$ of voxels in the domain $\mathbf{\Omega}\subseteq\mathbb{R}^{3}$, and the ground truth segmentation of heart structures as a mapping from intensity to the bi-ventricular myocardium region $f:\mathcal{I}\subseteq\mathbb{R}\mapsto\mathcal{S}$. Two dynUNets with the same structure are individually employed for CMR and CT segmentation. The dynUNets are structured with five down-size convolution layers, four up-size isotropic convolution layers, a bottleneck of 128 channels, and a size of $4\times4\times4$. Both take uncropped image volumes and infer segmentation in a sliding window manner, implemented using \texttt{sliding\_window\_inference} from MONAI. The CMR dynUNet performs slice-wise 2D inference and reconstructs the volume from stacked predictions, while the CT dynUNet ingests the full 3D volume and outputs a volumetric segmentation in one pass. The segmentation is downsampled to equalize the in-plane and through-plane resolutions. This approach mitigates the anisotropic nature of CMR data and minimizes resolution differences between CMR and CT data. 

We assume that cardiac anatomy is consistent between imaging modalities, but is observed differently in CMR and CT images. Unlike the CT image, the ventricular cardiac anatomy in the CMR image (in Fig \ref{fig:graphical_summary}) is incomplete due to incomplete or erroneous segmentations near the apex and basal plane. Moreover, aliasing appears due to the large slice-to-slice distance, and slice misalignments caused by motion artifacts are evident. The previous segmentation and downsampling approach mitigates aliasing and misalignments, but critical slices must be restored. Therefore, we used a ResNet to predict a complementary segmentation from dynUNet's last feature map output for segmentation completion. 

The segmentation completion combines dynUNet's segmentation with ResNet's complementary segmentation through distance-weighted blending. A sigmoid-transformed mask $\mathcal{W}$ is generated from the signed distance field of the dynUNet segmentation, creating smooth weighting coefficients that prioritize ResNet contributions near existing segmentation boundaries while excluding regions farther from the boundary and preserving anatomical continuity.
\begin{align}
    \mathcal{W} &=M_{\text{BG}}\odot\sigma(\alpha\cdot\mathcal{D} + 1) \\
    \hat{\mathcal{S}} &= \hat{\mathcal{S}}_{\text{+}} + \mathcal{W} \odot \hat{\mathcal{S}}_{\text{-}} \label{eq:segmentation_completion_algorithm}
\end{align}
where $\odot$ denotes pixel-wise product, $M_{\text{BG}}$ represents the background mask (0 for bi-ventricular myocardium region's pixels, 1 elsewhere) extracted from dynUNet's segmentation output, $\mathcal{D}$ denotes the signed distance field with positive values inside myocardium regions and negative values outside, scale factor $\alpha=0.86$ controls the contribution from ResNet's complementary segmentation $\hat{\mathcal{S}}_{\text{-}}$, and $\hat{\mathcal{S}}_{\text{+}}$ represents the dynUNet's segmentation.

Shown in Fig~\ref{fig:overview}, CT dynUNet and CMR dynUNet are optimized by minimizing the segmentation loss $\mathcal{L}_{\text{Seg}}$, which combines losses from both CT and CMR processing flows during training.
\begin{equation}
\begin{aligned}
    \mathcal{L}_{\mathrm{Seg}} &= \mathcal{L}_{\mathrm{Seg, CT}} + \mathcal{L}_{\mathrm{Seg, CMR}} \\
    \mathcal{L}_{\mathrm{Seg,\bullet}}(\hat{\mathcal{S}}_{\text{+}}, \mathcal{S}_{\text{+}}) &= 1 - \frac{2|\hat{\mathcal{S}}_{\text{+}} \cap \mathcal{S}_{\text{+}}|}{|\hat{\mathcal{S}}_{\text{+}}| + |\mathcal{S}_{\text{+}}|}
\end{aligned}
\end{equation}
where $\mathcal{L}_{\mathrm{Seg,\bullet}}$ represents the Dice loss for either CT or CMR processing, $\hat{\mathcal{S}}_{\text{+}}$ denotes dynUNet's segmentation, and $\mathcal{S}_{\text{+}}$ is the corresponding ground truth segmentation.

ResNet is trained solely on CT data using masked CT segmentations to learn restoration of basal and apical regions. Following equation \ref{eq:segmentation_completion_algorithm}, ResNet is optimized by minimizing a similar Dice loss $\mathcal{L}_{\text{SegC}}$, which is calculated between the complete segmentation prediction ($\hat{\mathcal{S}}$) and the ground truth ($\mathcal{S}$). The CT dynUNet is frozen and provides the partial segmentation as a result of masking of the basal and apical slices. This loss function is written as:
\begin{align}
    \mathcal{L}_{\mathrm{SegC}}(\hat{\mathcal{S}}, \mathcal{S}) &= 1 - \frac{2|\hat{\mathcal{S}} \cap \mathcal{S}|}{|\hat{\mathcal{S}}| + |\mathcal{S}|}
\end{align}
Eventually, the dynUNet's segmentation and the ResNet restoration are complementary and convey the underlying cardiac anatomy. The distance map results from a Euclidean distance transform applied to the complete myocardium segmentation, implemented using \texttt{distance\_transform\_edt} from MONAI.

\subsection{Adaptive Subdivision Process}
\label{sec:gsn}
Based on a subdivision shape modeling process elaborated in \cite{maugerIterativeDiffeomorphicAlgorithm2018}, we adapted a biventricular template mesh $\mathcal{M_{\text{c}}}=\left(\mathcal{V}, \mathcal{F}\right)$ for left and right ventricular myocardium, with 388 vertices ($\mathcal{V}=388$) and 780 faces ($\mathcal{F}=780$). Each vertex is assigned an anatomical label as one of LV endocardium, RV endocardium, LV epicardium, RV epicardium or valve. This mesh was defined in the Normalized Device Coordinates, where the mesh's size is bounded in the $\left[-1,1\right]$ range and centered at the origin of the space. 

\begin{figure}[!t]
    \centering
    \includegraphics[width=\columnwidth]{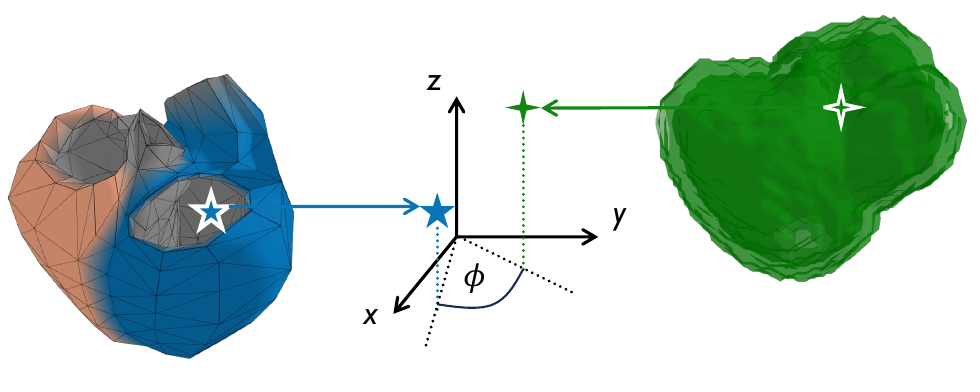}
    \\[1ex]
    \small (a) Step 1: Rigid ``swing'' rotation
    \\[2ex]
    \includegraphics[width=\columnwidth]{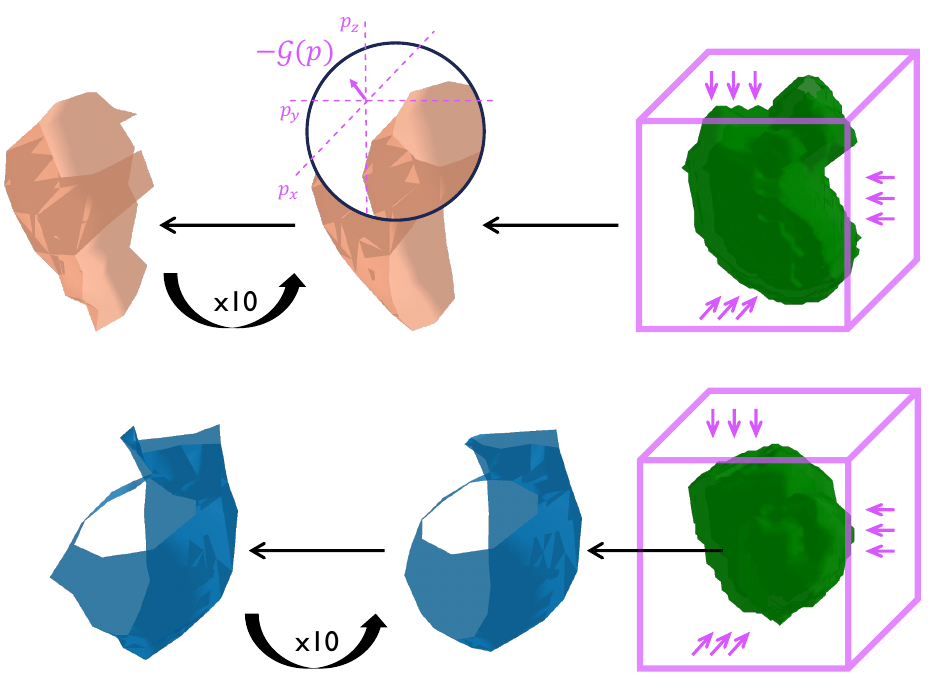}
    \\[1ex]
    \small (b) Step 2: Gradient field deformation
    \caption{\textbf{The patient-specific adjustment applied to the template model.} (a) A rigid ``swing'' rotation is determined to register the RV centroid in the template mesh with the RV centroid in the distance map (shown as extracted myocardium surface in the figures for ease of demonstration). (b) The gradient field is calculated from the distance map and is used to deform the template mesh iteratively along the reverse direction of the gradient vectors. The gradient field deformation is applied individually to groups of the template mesh's vertices categorized by their anatomical label. Only LV and RV epicardium are shown in the figures for ease of demonstration.}
    \label{fig:gradient_deform}
\end{figure}

Adjusting $\mathcal{M_{\text{c}}}$ in the gradient field follows a two-step procedure visualized in Fig \ref{fig:gradient_deform}. The first step is to align the RV centroid in $\mathcal{M_{\text{c}}}$ with the RV centroid in the distance map. This alignment is achieved through a rigid ``swing'' rotation applied to $\mathcal{M}_{\text{c}}$, where the RV centroid (extracted from its endocardial surface) serves as the reference point for the registration. After translating and scaling the distance map to Normalized Device Coordinates, the RV centroid in the distance map is identified and serves as the anchor point. Because all data were registered to the heart coordinate system in the data preprocessing, and the long axis of $\mathcal{M}_{\text{c}}$ is parallel to the long axis of the distance map and the global $Z$-axis, we rotate $\mathcal{M}_{\text{c}}$ about the $Z$-axis by an angle $\phi$. This angle is determined by the projection of the two RV centroids onto the $XY$-plane. Empirically, this rigid swing rotation yields the most reliable initial shape alignment. $\mathcal{M_{\text{c}}}$ and the distance map are rescaled to the original size of the distance map after aligning the RV centroid.

The second step is a gradient-field mesh deformation. In the distance map, we find the gradient field using the default configuration of PyTorch's \texttt{torch.gradient} function \cite{paszke2019pytorch}, which employs a hybrid approach: first-order estimates are used for boundary points where neighboring points may be missing on one side, while second-order estimates are applied for interior points where complete neighborhood information is available. This hybrid approach balances accuracy with computational feasibility at domain boundaries. Three partial derivatives $\left(\partial \bm{p}/\partial x,\partial \bm{p}/\partial y,\partial \bm{p}/\partial z\right)$ determine the gradient vector $\mathcal{G}\left(\bm{p}\right)$. Deforming the $\mathcal{M_{\text{c}}}$ is performed by moving its vertices in the gradient field in iterations. This results in a smooth trajectory that leads each vertex to the zero gradient position to match the underlying surface. Each vertex is iteratively moved in the reverse direction of the gradient vector $\mathcal{G}\left(\bm{p}\right)$ by a controlled distance:
\begin{equation}
    d = 
    \begin{cases}
        1 & \text{if } \left|\mathcal{G}\left(\bm{p}\right)\right|_2 > 1 \\
        \left|\mathcal{G}\left(\bm{p}\right)\right|_2 & \text{if } \left|\mathcal{G}\left(\bm{p}\right)\right|_2\leq1
    \end{cases}
\end{equation}
This controlled step size ensures smooth translation and diffeomorphic deformation, preserving both quality and topology of the template mesh. Theoretically, convergence is achieved when $\left|\mathcal{G}\left(\bm{p}\right)\right|_2=0$ for each vertex. We conducted preliminary experiments to identify that 10 iterations consistently produced the smallest average surface distance error between the ground-truth surface (extracted from the zero-level set of the distance map) and the deformed template mesh surface while maintaining reasonable computational cost, with minimal variation in convergence behavior between cases. This process is performed on vertices with the same anatomical label and individually for each anatomical label group. 

Our \textbf{GSN} structure, inspired by the Loop subdivision \cite{loop1987smooth}, refines the mesh by splitting edges to form new triangles and updating the vertices to approximate intricate heart geometries. In Loop subdivision, the position of the updated vertex $\bm{v}^{\prime}$ corresponding to original vertex $\bm{v}_{i}$ is determined by a weighted combination of $\bm{v}_{i}$ and its neighbors $\mathcal{N}(\bm{v}_{i})$ as follows:
\begin{align}
    \bm{v}^{\prime}_{i}&=(1-\alpha\cdot\text{deg}(i))\bm{v}_{i}+\alpha\sum_{\bm{v}_j\in\mathcal{N}(\bm{v}_{i})}\bm{v}_{j} \notag \\
    &=\bm{v}_{i}+\sum_{\bm{v}_{j}\in\mathcal{N}(\bm{v}_{i})}\alpha(\bm{v}_{j}-\bm{v}_{i}) \label{eq:vert_reposition} \\
    \text{deg}(i)&=\sum_{j}\mathbf{A}_{i,j}+\mathbf{I}_{i,i}
    \label{eq:deg}
\end{align}
where $\text{deg}(i)$ is the vertex's degree, $\mathbf{A}\in\mathbb{R}^{N\times N}$ is the adjacency matrix of the triangular mesh with $N$ vertices, and $\mathbf{I}\in\mathbb{R}^{N\times N}$ is an identity matrix added to stabilize the subdivision process, and $\alpha$ is a weight defined by Warren's formula \cite{zorin1996interpolating}. Splitting edges results in new vertices --- the mid-points of those edges. 

In this study, we avoid mesh shrinkage and provide greater deformation freedom in our GSN structure. GSN updates both original and new vertices following equations in Loop subdivision but substituting $\alpha$ with a three-layer, 16 hidden-feature optimizable multi-layer perceptron $h_{\theta_{\text{m}}}$ and normalizing the weighted sum by vertex's degree. This update to all vertices is formalized as 
\begin{equation}
    \mathrm{\Delta}\bm{v}_{i}=\sum_{\bm{v}_{j}\in\mathcal{N}(\bm{v}_{i})}\frac{1}{\sqrt{\text{deg}(i)}\cdot\sqrt{\text{deg}(j)}}\cdot h_{\theta_{\text{m}}}\left(\bm{v}_{j}-\bm{v}_{i}\right)
    \label{eq:gsn}
\end{equation}
where the normalization factor $1/(\sqrt{\text{deg}(i)}\cdot\sqrt{\text{deg}(j)})$ serves to improve numerical stability during training, inspired by spectral graph theory as formalized in \cite{kipf2016semi}, where similar normalization prevents gradient scaling issues in graph neural networks.

After many subdivision levels, the new faces remain triangular and new vertices inherit the anatomical label, ensuring fixed mesh topology. Following our subdivision paradigm and starting with the same template mesh, a dense point correspondence linking every point from one mesh to its equivalent point on the other mesh is also ensured.

We optimized the GSN by minimizing the Chamfer distance \cite{fan2017point} between the generated dense 3D mesh model $\hat{\mathcal{M}}=\left(\hat{\mathcal{V}},\hat{\mathcal{F}}\right)$ and the ground truth point clouds. These point clouds $\mathcal{P}^{\prime}$ are extracted from ground truth segmentation. The Chamfer distance $\mathcal{L}_{\text{Chmf}}(\hat{\mathcal{M}},\mathcal{P}^{\prime})$ is minimized along with the Laplacian smoothing \cite{desbrun1999implicit} $\mathcal{L}_{\text{Lap}}(\hat{\mathcal{M}})$,
\begin{align}
    \mathcal{L}_{\text{Chmf}}(\hat{\mathcal{M}},\mathcal{P}^{\prime}) &= \sum_{i=0}^{N}{\frac{1}{\left|\hat{\mathcal{V}_{i}}\right|} \sum_{v \in \hat{\mathcal{V}}_{i}}{\min_{p \in \mathcal{P}^{\prime}_{i}} {\left|v - p\right|}_{2}^{2}}}  \notag \\ 
    \label{eq:gsn_loss0}
    &+ \sum_{i=0}^{N}{\frac{1}{\left|\mathcal{P}^{\prime}_{i}\right|} \sum_{p \in \mathcal{P}^{\prime}_{i}}{\min_{v \in \hat{\mathcal{V}}_{i}} {\left|p - v\right|}_{2}^{2}}} \\
    \mathcal{L}_{\text{Lap}}(\hat{\mathcal{M}}) &= \sum_{v_i \in \hat{\mathcal{V}}}\frac{\left(\cot{a}_{ij} + \cot{b}_{ij}\right)}{4A_i}\left(v_i - v_j\right)
    \label{eq:gsn_loss1}
\end{align}
where $v$ is vertices with the i-th anatomical label in $\hat{\mathcal{M}}$ and $p$ is the points' position in $\mathcal{P}^{\prime}$ extracted from the ground truth segmentation corresponding to that anatomical label; ${a}_{ij}$ and ${b}_{ij}$ are the ``outside'' angles in the two triangles connecting vertex $v_i$ and its neighboring vertices $v_j$, and $A_i$ is the sum of the areas of all triangles containing vertex $v_i$, implemented using \texttt{chamfer\_distance} and \texttt{mesh\_laplacian\_smoothing} from PyTorch3D \cite{ravi2020accelerating}.

\subsection{Optimization}
We empirically apply weight coefficients $\mathrm{\lambda}_{0}=0.66$ and $\mathrm{\lambda}_{1}=0.75$ to balance the loss components in the total loss for training, given by
\begin{equation}
\mathcal{L}=\mathcal{L}_{\text{Seg}}+\mathcal{L}_{\text{SegC}}+\mathrm{\lambda}_{0}\cdot\mathcal{L}_{\text{Chmf}}+\mathrm{\lambda}_{1}\cdot\mathcal{L}_{\text{Lap}}
 \label{eq:total_loss}
\end{equation}

MorphiNet comprises three separate modules with learnable parameters: dynUNets, ResNet, and GSN, trained sequentially. Equation \ref{eq:total_loss} represents the total loss function that guides the overall training process, with different components becoming active during the training of specific modules: $\mathcal{L}_{\text{Seg}}$ is active during dynUNet training (synchronize training with CMR dynUNet and CT dynUNet for 100 epochs), $\mathcal{L}_{\text{SegC}}$ is active during ResNet training (80 epochs), and $\mathcal{L}_{\text{Chmf}}$ and $\mathcal{L}_{\text{Lap}}$ are active during GSN training (120 epochs). 

\section{Experiment}
\subsection{Experimental Setup}
For training, we used a subset of 232 CT scans randomly chosen from the Scottish Computed Tomography of the Heart (SCOT-HEART) dataset \cite{scot2018coronary}, and a subset of 222 CMR scans (111 cases at ED and ES) randomly chosen from the Cardiac Atlas Project (CAP) dataset \cite{govil2023deep} consisting of patients with tetralogy of Fallot. For testing, hold out splits from SCOT-HEART and CAP, in addition to two external test sets --- the Multimodal Whole Heart Segmentation (MMWHS) with 20 CT scans \cite{zhuang2018multivariate} and a subset of 20 CMR scans randomly chosen from the ACDC \cite{bernard2018deep}. The CT and CMR images in our dataset are unpaired. After randomization, an 80-20 training/testing split (by participant) was applied to the CAP and SCOT-HEART datasets, respectively. Except for the CMR dynUNet input in MorphiNet, where images and manual segmentations are kept in their original resolution, all images were interpolated by bilinear interpolation, and all manual segmentations were interpolated by nearest-neighbor interpolation. Both resulted in a resolution of $2\ \text{mm}\times2\ \text{mm}\times2\ \text{mm}$. All data and template mesh were registered to the heart coordinate system. The data augmentation included Gaussian noise with a standard deviation of 0.01, Gaussian smoothing with an isotropic kernel of 0.25 mean and 1.5 standard deviations, and intensity scaling with a factor of 1.3 --- all operations were applied with 0.15 probability to individual data.

MorphiNet was trained using the AdamW optimizer \cite{loshchilov2017decoupled} with an initial learning rate of 0.001 for 300 epochs: 100 for dynUNets, 80 for ResNet, and 120 for GSN. Training and testing were conducted on an NVIDIA RTX 3090 GPU, the same as all baseline methods. We determined the hyperparameters via sensitivity analysis and describe the training process in detail in the supplementary material. For comparison, we selected representative baselines with publicly available implementations that are tailored to cardiac or closely related anatomical shape reconstruction and that span complementary state-of-the-art modeling paradigms (template-based methods and implicit neural functions). To do this, we implemented five baseline methods: Voxel2Mesh \cite{wickramasinghe2020voxel2mesh}, ModusGraph \cite{deng2023modusgraph}, HeartDeformNet \cite{kong2022learn}, CorticalFlow++ \cite{santa2022corticalflow++} and the NDF \cite{sun2022topology}. \textbf{Voxel2Mesh} predicts meshes via cascaded template deformation and unpooling; \textbf{CorticalFlow++} relies on diffeomorphic template flows originally designed for cortical surfaces; \textbf{ModusGraph} predicts deformations of a whole-heart template mesh; \textbf{HeartDeformNet} leverages biharmonic-coordinate control handles to drive voxel-guided template deformation; and \textbf{NDF} provides an implicit, topology-preserving shape representation. All methods used identical training/testing datasets and preprocessing and post-processing steps for fair comparison. 

A Laplacian smoothing filter $(\lambda=0.13)$ is applied to all methods' generated models to achieve preferable surface smoothness. Methods \textbf{except MorphiNet} require a region of interest cropping, resizing, and zeros padding to $128^{3}$ pixel-size volumes. Methods \textbf{except NDF} use the identical template mesh described in section \ref{sec:distance_map}. NDF operates on ground-truth segmentation masks and does not provide a learned segmentation component: its meshes were obtained by extracting the zero-level set from the signed distance representation of the reference segmentation using marching cubes, followed by a surface decimation to control the number of vertices and faces. Therefore, NDF is a ``high benchmark’’ in performance metric comparisons. Although marching cubes result in high accuracy measured by the performance metrics in Table \ref{table:ct_result}, the resulting meshes cannot be used for digital heart computational analysis due to the lack of correspondence between cases. 

We reimplemented HeartDeformNet~\cite{kong2022learn} based on the official repository \texttt{https://github.com/fkong7/HeartDeformNets}, incorporating modifications to ensure compatibility with modern GPU architectures and enable flexible template mesh selection. This implementation is available at \texttt{https://github.com/MalikTeng/HeartDeformNets}. We trained and evaluated this version using our standard experimental setup, reporting the metric scores as \textbf{HeartDeformNet*} for both CT and CMR reconstructions. Additionally, we evaluated the original pre-trained networks and template meshes provided by Kong \emph{et al.}, denoted as \textbf{HeartDeformNet**}. Notably, both implementations failed to produce robust reconstructions for the thin RV myocardium, likely due to numerical limitations inherent in the biharmonic coordinate calculation. Consequently, we restricted the evaluation of these models to the LV myocardium. To facilitate a fair comparison, we retrained MorphiNet using an updated LV-only template mesh. NDF results were similarly updated to serve as a ``target benchmark'' for LV surface reconstruction.

\subsection{Surface Reconstruction Performance}
\label{sec:surface_reconstruction_performance}
We first evaluated MorphiNet's reconstruction capability on both CT and CMR datasets. Dice score (\textbf{Dice}) and Hausdorff distance (\textbf{Hd}) \cite{taha2015metrics} were used to quantify errors between voxelized, generated models and ground truth segmentations, using the implementation from \url{https://github.com/cvlab-epfl/voxel2mesh}. All voxelized generated models and ground truth segmentations were resampled to $128^{3}$ voxel grids, and Dice/Hd were computed on the bi-ventricular myocardium (LV+RV myocardium). GPU Inference Time (\textbf{InFt}) is reported as the average time to generate a myocardial mesh from CT/CMR images. To assess whether the reconstructed surfaces are suitable for downstream digital-heart analysis, we further report the following mesh-based metrics on the extracted bi-ventricular myocardial surfaces:

\subsubsection{Average Surface Distance (\textbf{ASD}) \cite{kim2012bidirectional}} 
ASD measures the mean minimum Euclidean distance between the generated model surface and a set of 5{,}000 points uniformly sampled from the ground truth myocardial surface, where 5{,}000 approximates the average number of vertices on the generated meshes. Lower ASD indicates closer geometric agreement between predicted and reference myocardium.

\subsubsection{Aspect Ratio (\textbf{AspR})}
AspR characterizes the proportionality of element dimensions for each triangular face, computed from the edge lengths such that equilateral triangles have $\text{AspR} \approx 1.0$, while increasingly stretched or skewed elements yield larger values. We report the mean AspR over all myocardial surface elements, where values below 5 are typically considered acceptable. AspR is computed using \texttt{compute\_cell\_quality} in \texttt{pyvista}. 

\subsubsection{Scaled Jacobian Ratio (\textbf{JacR})}
JacR evaluates element shape quality by comparing the Jacobian of the mapping from a reference element to each mesh element against an ideal configuration. The scaled Jacobian lies in the range $[-1, 1]$, where $1.0$ denotes an undistorted, well-shaped element, values near $0$ indicate highly distorted elements, and negative values imply inverted or invalid elements. We report the mean JacR over all myocardial elements, computed with \texttt{compute\_cell\_quality} in \texttt{pyvista}. 

\subsubsection{Mean Normal Consistency (\textbf{MnC})}
MnC measures surface normal ($\mathbf{n}$) alignment quality as
\[
\text{MnC} = 0.5 \times \bigl(\text{NC}_{\text{pred}\rightarrow\text{gt}} + \text{NC}_{\text{gt}\rightarrow\text{pred}}\bigr),
\]
where $\text{NC}_{\text{pred}\rightarrow\text{gt}} = \text{mean}\bigl(|\cos(\mathbf{n}_{\text{pred}}, \mathbf{n}_{\text{gt\_nearest}})|\bigr)$ over sampled myocardial surface points, and analogously for $\text{NC}_{\text{gt}\rightarrow\text{pred}}$ \cite{sun2022topology}. MnC ranges from 0 to 1, with higher values indicating smoother, more coherent surface normals.

\subsubsection{Non-manifold Faces Ratio (\textbf{NmF})}
NmF quantifies geometric and topological defects as the proportion of adjacent face pairs with opposing normal directions, defined as
\[
\text{NmF} = \frac{\text{count}\bigl(\mathbf{n}_i \cdot \mathbf{n}_j < 0\bigr)}{\text{count(total faces)}} ,
\]
for adjacent faces $i$ and $j$ on the myocardial surface, where lower values indicate better manifold properties and surface regularity \cite{sun2022topology}. 

\begin{figure*}[!t]
    \centering
    \includegraphics[width=\textwidth]{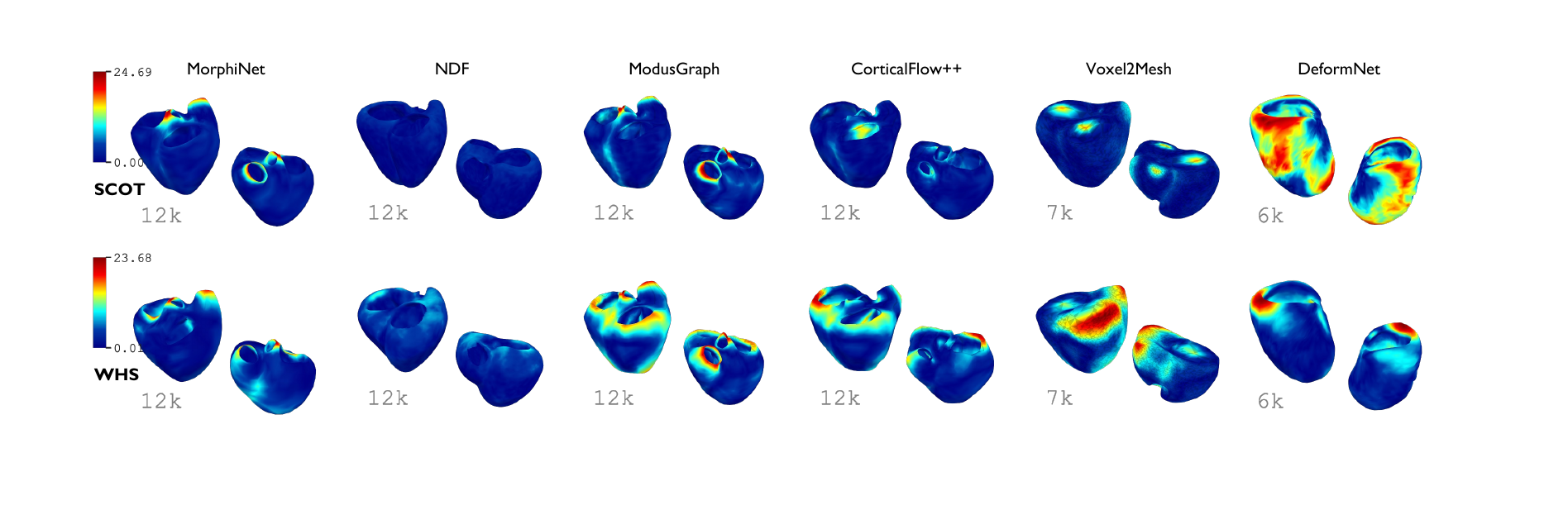}
    \caption{\textbf{Surface reconstruction results on SCOT-HEART and MMWHS CT datasets}, with the highest face count per method displayed in the bottom left and colored with surface errors (mm). The RV myocardium segmentation in the MMWHS dataset grows from RV epicardium, resulting in a 3 mm-thick myocardium free-wall segmentation.}
    \label{fig:recon_result_ct}
\end{figure*}

Fig.~\ref{fig:recon_result_ct} visualizes the surface reconstruction results on the \textbf{SCOT}-HEART and MM\textbf{WHS} datasets, with quantitative comparisons detailed in Table~\ref{table:ct_result}. The CT reconstruction task requires balancing high geometric accuracy --- minimizing the error between the voxelized model and the ground truth segmentation --- with the maintenance of high-quality mesh topology, including fixed connectivity and element regularity. Among template-based methods, \textbf{MorphiNet} achieves superior anatomical fidelity, giving a Dice score of 0.8 and a Hd of 1.5 on the SCOT dataset, significantly exceeding ModusGraph (Dice 0.4, Hd 6.1). In the specific assessment of LV myocardium reconstruction, MorphiNet (Dice 0.9, Hd 1.4) shows substantial improvements over both reimplemented HeartDeformNet$\star$ (Dice 0.4) and original pre-trained HeartDeformNet$\star\star$ (Dice 0.7). This confirms that our gradient-field deformation provides more precise geometrical conformance to patient-specific anatomy than the biharmonic coordinate control-handle deformation employed by HeartDeformNet. Although the neural implicit function (NDF) attains a marginally lower ASD (1.0mm) through pixel-wise optimization, it incurs a heavy computational cost (133.1s). In contrast, MorphiNet delivers comparable accuracy with high efficiency (2.7s) while ensuring the dense point correspondence and fixed topology inherent to template-based methods. We address the observed trade-off between this surface refinement and element regularity (AspR 1.6, JacR 0.6) in Section~\ref{sec:mesh_quality_worsening}.

\begin{table*}[!t]
    \centering
    \caption{\textbf{Quantitative evaluation of surface reconstruction on SCOT-HEART and MMWHS CT datasets}, highlighting MorphiNet's superior anatomical accuracy among template-based methods. Arrows ($\uparrow$/$\downarrow$) indicate the direction of the ideal value (e.g., higher Dice/JacR/MnC is better; lower Hd/ASD/AspR/NmF/InFt is better). Statistical significance is assessed via one-way ANOVA ($p<0.001$) and Tukey's HSD test. Superscript letters denote Compact Letter Display (CLD) groupings: methods sharing a letter are statistically indistinguishable ($p \ge 0.05$), with `a' representing the best performance. The best scores of \textsuperscript{$\diamond$} template-based methods are in bold, and the \textsuperscript{$\dagger$} neural implicit function gives high target benchmark scores.}
    \label{table:ct_result}
    \vspace{0.2cm}
    \begin{tabularx}{\textwidth}{c c | Y Y Y Y Y Y Y c}
    \hline
        \textbf{Dataset} & \textbf{Method} & \textbf{Dice}$\uparrow$ & \textbf{Hd}$\downarrow$ & \textbf{ASD}$\downarrow$ & \textbf{AspR}$\downarrow$ & \textbf{JacR}$\uparrow$ & \textbf{MnC}$\uparrow$ & \textbf{NmF}$\downarrow$ & \textbf{InFt}$\downarrow$ \\
    \hline
        \addlinespace[4pt]
        & {\textsuperscript{$\diamond$}Voxel2Mesh} & \textsuperscript{c}$0.4\left(0.04\right)$ & \textsuperscript{b}$5.4\left(0.91\right)$ & \textsuperscript{e}$4.6\left(0.66\right)$ & \textsuperscript{a}$\textbf{1.4}\left(\textbf{0.01}\right)$ & \textsuperscript{b}$0.7\left(0.01\right)$ & \textsuperscript{d}$0.7\left(0.02\right)$ & \textsuperscript{c}$1.0\left(0.01\right)$ & \textsuperscript{a}$\textbf{0.3}\left(\textbf{0.03}\right)$ \\
        & {\textsuperscript{$\diamond$}CortFlow++} & \textsuperscript{d}$0.3\left(0.06\right)$ & \textsuperscript{d}$7.7\left(1.25\right)$ & \textsuperscript{c}$3.4\left(0.56\right)$ & \textsuperscript{c}$1.5\left(0.01\right)$ & \textsuperscript{c}$0.7\left(0.01\right)$ & \textsuperscript{c}$0.8\left(0.01\right)$ & \textsuperscript{b}$\textbf{0.0}\left(\textbf{0.00}\right)$ & \textsuperscript{a}$0.8\left(0.01\right)$ \\ 
    \textbf{SCOT} & {\textsuperscript{$\diamond$}ModusGraph} & \textsuperscript{b}$0.4\left(0.06\right)$ & \textsuperscript{c}$6.1\left(1.23\right)$ & \textsuperscript{d}$3.8\left(0.68\right)$ & \textsuperscript{b}$1.4\left(0.00\right)$ & \textsuperscript{a}$\textbf{0.7}\left(\textbf{0.00}\right)$ & \textsuperscript{c}$0.8\left(0.01\right)$ & \textsuperscript{b}$\textbf{0.0}\left(\textbf{0.00}\right)$ & \textsuperscript{a}$0.7\left(0.02\right)$ \\
        & {\textsuperscript{$\diamond$}\textbf{MorphiNet}} & \textsuperscript{a}$\textbf{0.8}\left(\textbf{0.02}\right)$ & \textsuperscript{a}$\textbf{1.5}\left(\textbf{0.20}\right)$ & \textsuperscript{b}$\textbf{1.4}\left(\textbf{0.75}\right)$ & \textsuperscript{d}$1.6\left(0.04\right)$ & \textsuperscript{d}$0.6\left(0.02\right)$ & \textsuperscript{b}$\textbf{0.8}\left(\textbf{0.01}\right)$ & \textsuperscript{b}$0.0\left(0.01\right)$ & \textsuperscript{a}$2.7\left(0.18\right)$ \\
        & {\textsuperscript{$\dagger$}NDF} & \textsuperscript{a}\textit{0.8(0.06)} & \textsuperscript{a}\textit{1.7(0.63)} & \textsuperscript{a}\textit{1.0(0.25)} & \textsuperscript{e}\textit{1.7(0.01)} & \textsuperscript{e}\textit{0.6(0.01)} & \textsuperscript{a}\textit{0.8(0.01)} & \textsuperscript{a}\textit{0.0(0.00)} & \textsuperscript{b}\textit{133.1(39.9)} \\

    \hdashline
        \addlinespace[4pt]
        & {\textsuperscript{$\diamond$}HeartDeformNet$\star$} & \textsuperscript{c}$0.4\left(0.13\right)$ & \textsuperscript{d}$7.8\left(1.88\right)$ & \textsuperscript{c}$3.6\left(1.03\right)$ & \textsuperscript{c}$1.8\left(0.23\right)$ & \textsuperscript{c}$0.5\left(0.07\right)$ & \textsuperscript{a}$0.8\left(0.10\right)$ & \textsuperscript{b}$0.0\left(0.00\right)$ & \textsuperscript{a}$\textbf{0.5}\left(\textbf{0.06}\right)$ \\
        & {\textsuperscript{$\diamond$}HeartDeformNet$\star\star$} & \textsuperscript{b}$0.7\left(0.10\right)$ & \textsuperscript{c}$4.7\left(0.91\right)$ & \textsuperscript{c}$3.6\left(0.92\right)$ & \textsuperscript{a}$\textbf{1.5}\left(\textbf{0.19}\right)$ & \textsuperscript{a}$\textbf{0.6}\left(\textbf{0.08}\right)$ & \textsuperscript{a}$0.8\left(0.10\right)$ & \textsuperscript{a}$\textbf{0.0}\left(\textbf{0.00}\right)$ & \textsuperscript{c}$13.3\left(2.8\right)$ \\
    \textbf{SCOT (LEFT)} & {\textsuperscript{$\diamond$}\textbf{MorphiNet}} & \textsuperscript{a}$\textbf{0.9}\left(\textbf{0.11}\right)$ & \textsuperscript{a}$\textbf{1.4}\left(\textbf{0.26}\right)$ & \textsuperscript{a}$\textbf{2.0}\left(\textbf{0.80}\right)$ & \textsuperscript{b}$1.6\left(0.21\right)$ & \textsuperscript{b}$0.6\left(0.08\right)$ & \textsuperscript{a}$\textbf{0.8}\left(\textbf{0.10}\right)$ & \textsuperscript{c}$0.0\left(0.01\right)$ & \textsuperscript{b}$2.9\left(0.38\right)$ \\
        & {\textsuperscript{$\dagger$}NDF} & \textsuperscript{b}\textit{0.7(0.11)} & \textsuperscript{b}\textit{3.7(1.08)} & \textsuperscript{b}\textit{3.1(1.39)} & \textsuperscript{b}\textit{1.6(0.21)} & \textsuperscript{b}\textit{0.6(0.08)} & \textsuperscript{a}\textit{0.8(0.11)} & \textsuperscript{b}\textit{0.0(0.00)} & \textsuperscript{d}\textit{32.0(4.2)} \\
    \hline
        \addlinespace[4pt]
        & {\textsuperscript{$\diamond$}Voxel2Mesh} & \textsuperscript{c}$0.4\left(0.09\right)$ & \textsuperscript{b}$7.1\left(1.21\right)$ & \textsuperscript{d}$6.1\left(1.75\right)$ & \textsuperscript{a}$\textbf{1.4}\left(\textbf{0.03}\right)$ & \textsuperscript{b}$0.7\left(0.01\right)$ & \textsuperscript{c}$0.7\left(0.02\right)$ & \textsuperscript{d}$1.0\left(0.01\right)$ & \textsuperscript{a}$\textbf{0.4}\left(\textbf{0.01}\right)$ \\
        & {\textsuperscript{$\diamond$}CortFlow++} & \textsuperscript{d}$0.3\left(0.05\right)$ & \textsuperscript{c}$9.3\left(2.34\right)$ & \textsuperscript{c}$4.7\left(1.16\right)$ & \textsuperscript{a}$1.5\left(0.01\right)$ & \textsuperscript{c}$0.7\left(0.01\right)$ & \textsuperscript{c}$0.7\left(0.03\right)$ & \textsuperscript{bc}$\textbf{0.0}\left(\textbf{0.00}\right)$ & \textsuperscript{a}$0.8\left(0.01\right)$ \\
    \textbf{WHS} & {\textsuperscript{$\diamond$}ModusGraph} & \textsuperscript{c}$0.4\left(0.08\right)$ & \textsuperscript{b}$7.9\left(1.38\right)$ & \textsuperscript{cd}$5.2\left(1.37\right)$ & \textsuperscript{a}$1.4\left(0.00\right)$ & \textsuperscript{a}$\textbf{0.8}\left(\textbf{0.00}\right)$ & \textsuperscript{c}$0.7\left(0.03\right)$ & \textsuperscript{b}$\textbf{0.0}\left(\textbf{0.00}\right)$ & \textsuperscript{a}$0.6\left(0.01\right)$ \\
        & {\textsuperscript{$\diamond$}\textbf{MorphiNet}} & \textsuperscript{b}$\textbf{0.8}\left(\textbf{0.03}\right)$ & \textsuperscript{a}$\textbf{2.8}\left(\textbf{0.77}\right)$ & \textsuperscript{b}$\textbf{2.5}\left(\textbf{1.08}\right)$ & \textsuperscript{c}$1.8\left(0.25\right)$ & \textsuperscript{e}$0.6\left(0.07\right)$ & \textsuperscript{b}$\textbf{0.8}\left(\textbf{0.03}\right)$ & \textsuperscript{c}$0.0\left(0.02\right)$ & \textsuperscript{a}$2.4\left(0.12\right)$ \\
        & {\textsuperscript{$\dagger$}NDF} & \textsuperscript{a}\textit{0.8(0.04)} & \textsuperscript{a}\textit{1.6(0.27)} & \textsuperscript{a}\textit{1.2(0.21)} & \textsuperscript{b}\textit{1.6(0.01)} & \textsuperscript{d}\textit{0.6(0.01)} & \textsuperscript{a}\textit{0.8(0.01)} & \textsuperscript{a}\textit{0.0(0.00)} & \textsuperscript{b}\textit{135.2(26.7)} \\

    \hdashline
        \addlinespace[4pt]
        & {\textsuperscript{$\diamond$}HeartDeformNet$\star$} & \textsuperscript{c}$0.4\left(0.13\right)$ & \textsuperscript{c}$7.3\left(1.97\right)$ & \textsuperscript{c}$4.9\left(1.72\right)$ & \textsuperscript{a}$1.7\left(0.38\right)$ & \textsuperscript{a}$0.5\left(0.11\right)$ & \textsuperscript{a}$0.8\left(0.17\right)$ & \textsuperscript{a}$\textbf{0.0}\left(\textbf{0.00}\right)$ & \textsuperscript{a}$\textbf{0.5}\left(\textbf{0.10}\right)$ \\
        & {\textsuperscript{$\diamond$}HeartDeformNet$\star\star$} & \textsuperscript{b}$0.7\left(0.15\right)$ & \textsuperscript{b}$5.1\left(1.28\right)$ & \textsuperscript{bc}$4.0\left(1.06\right)$ & \textsuperscript{a}$\textbf{1.5}\left(\textbf{0.32}\right)$ & \textsuperscript{a}$\textbf{0.6}\left(\textbf{0.13}\right)$ & \textsuperscript{a}$0.8\left(0.17\right)$ & \textsuperscript{a}$\textbf{0.0}\left(\textbf{0.00}\right)$ & \textsuperscript{c}$7.3\left(1.7\right)$ \\
    \textbf{WHS (LEFT)} & {\textsuperscript{$\diamond$}MorphiNet} & \textsuperscript{ab}$\textbf{0.8}\left(\textbf{0.17}\right)$ & \textsuperscript{a}$\textbf{2.4}\left(\textbf{0.69}\right)$ & \textsuperscript{b}$\textbf{3.3}\left(\textbf{0.98}\right)$ & \textsuperscript{a}$1.7\left(0.37\right)$ & \textsuperscript{a}$0.5\left(0.12\right)$ & \textsuperscript{a}$\textbf{0.8}\left(\textbf{0.17}\right)$ & \textsuperscript{b}$0.0\left(0.01\right)$ & \textsuperscript{b}$2.9\left(0.63\right)$ \\
        & {\textsuperscript{$\dagger$}NDF} & \textsuperscript{a}\textit{0.8(0.18)} & \textsuperscript{a}\textit{2.5(1.32)} & \textsuperscript{a}\textit{1.6(0.86)} & \textsuperscript{a}\textit{1.5(0.34)} & \textsuperscript{a}\textit{0.6(0.13)} & \textsuperscript{a}\textit{0.8(0.18)} & \textsuperscript{a}\textit{0.0(0.00)} & \textsuperscript{d}\textit{20.0(4.4)} \\
    \hline
    \end{tabularx}
    \vspace{2mm}
    \begin{minipage}{\textwidth}
    \footnotesize
    \textit{Notes:} Dice and Hd are calculated based on pixels. ASD is in mm. AspR, JacR, MnC, and NmF are dimensionless. InFt is in seconds. SCOT (LEFT) and WHS (LEFT) indicate evaluation done solely with left-ventricular myocardium. Numbers represent average values and (standard deviation). 
    \end{minipage}
\end{table*}

\begin{figure*}[!t]
\includegraphics[width=\textwidth]{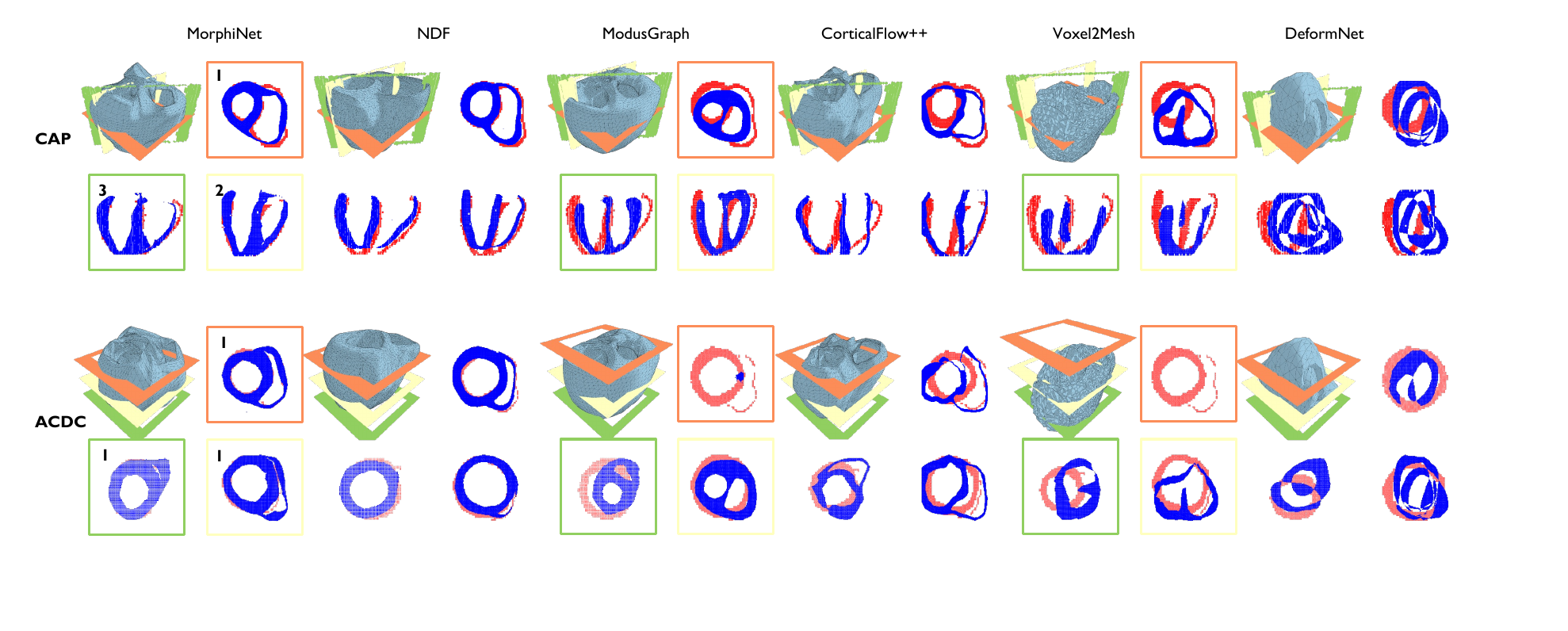}
\caption{\textbf{Surface reconstruction results on the CAP and ACDC CMR datasets}, depicting myocardium mesh slices in three views: 1) short-axis, 2) three-chamber, and 3) four-chamber. Note that the ACDC dataset provides no LAX image-view data. Those image-view planes are labeled as color boxes and positioned against respective reconstructed models. In each boxed image view, the model's cross-section (\textcolor{blue}{blue}) is visualized in contrast with the manual segmentation (\textcolor{red}{red}). The RV myocardium segmentation in the ACDC dataset grows from RV epicardium, resulting in a 3 mm-thick myocardium free-wall segmentation.}
\label{fig:recon_result_mr}
\end{figure*}

For CMR reconstruction, evaluated on the \textbf{CAP} and \textbf{ACDC} datasets, we present visualizations in Fig.~\ref{fig:recon_result_mr} and quantitative metrics in Table~\ref{table:mr_result}. Given that all methods receive only short-axis (SAX) stacks as input, we establish the long-axis (LAX) view metrics (Dice and Hd) as the critical indicator of \textit{anatomical fidelity}, the ability to hallucinate the unobserved longitudinal profile and missing apical/basal slices. MorphiNet demonstrates superior fidelity in this regard, achieving a LAX Dice of 0.6 and Hd of 7.1 pixels. This stands in stark contrast to other template-based methods like ModusGraph and CorticalFlow++, which struggle to infer the latent 3D geometry (LAX Dice 0.3, Hd $>10$), and the reimplemented HeartDeformNet$\star$, which fails to bridge the large inter-slice gaps of the LV (Hd 14.1). Notably, MorphiNet beats the anatomical completeness of the NDF benchmark (LAX Dice 0.5) despite NDF having the advantage of optimizing directly against the ground truth segmentation. While NDF achieves a marginally lower ASD (2.1 mm) through this direct fitting, it lacks point correspondence and incurs a prohibitive computational cost (138.7 s). MorphiNet leverages its learned gradient fields to deliver comparable anatomical realism and precise topology in just 2.7 s, with robust zero-shot generalization to the ACDC dataset (Dice 0.7).

\begin{table*}[!t]
    \centering
    \caption{\textbf{Quantitative evaluation of surface reconstruction on CAP and ACDC CMR datasets}, demonstrating MorphiNet's superior anatomical accuracy (highest $\uparrow$Dice, lowest $\downarrow$Hd/$\downarrow$ASD) among template-based methods with clinically feasible inference times ($\downarrow$InFt). Arrows indicate the optimal direction for each metric ($\uparrow$ higher is better; $\downarrow$ lower is better), with statistical significance assessed via one-way ANOVA ($p<0.001$) and Tukey's HSD test. Superscript letters denote Compact Letter Display (CLD) groupings: methods sharing a letter are statistically indistinguishable ($p \ge 0.05$), with `a' representing the best performance. The best scores of \textsuperscript{$\diamond$} template-based methods are in bold. The \textsuperscript{$\dagger$} neural implicit function gives high target benchmark scores.}
    \label{table:mr_result}
    \begin{tabularx}{\textwidth}{c c | Y Y Y Y Y Y Y Y c}
    \hline
        \textbf{Dataset} & \textbf{Method} & \textbf{Dice}{\tiny SAX}$\uparrow$ & \textbf{Dice}{\tiny LAX}$\uparrow$ & \textbf{Hd}{\tiny SAX}$\downarrow$ & \textbf{Hd}{\tiny LAX}$\downarrow$ & \textbf{ASD}$\downarrow$ & \textbf{AspR}$\downarrow$ & \textbf{JacR}$\uparrow$ & \textbf{NmF}$\downarrow$ & \textbf{InFt}$\downarrow$ \\
    \hline
        \addlinespace[4pt]
        & {\textsuperscript{$\diamond$}Voxel2Mesh} & \textsuperscript{c}$0.4\left(0.09\right)$ & \textsuperscript{c}$0.3\left(0.11\right)$ & \textsuperscript{d}$11.0\left(3.26\right)$ & \textsuperscript{b}$14.6\left(6.82\right)$ & \textsuperscript{d}$5.8\left(1.76\right)$ & \textsuperscript{b}$1.4\left(0.03\right)$ & \textsuperscript{b}$0.7\left(0.02\right)$ & \textsuperscript{d}$1.0\left(0.01\right)$ & \textsuperscript{a}$\textbf{0.3}\left(\textbf{0.01}\right)$ \\
        & {\textsuperscript{$\diamond$}CortFlow++} & \textsuperscript{d}$0.3\left(0.07\right)$ & \textsuperscript{c}$0.3\left(0.09\right)$ & \textsuperscript{c}$9.0\left(2.06\right)$ & \textsuperscript{b}$12.5\left(5.71\right)$ & \textsuperscript{b}$\textbf{3.8}\left(\textbf{1.10}\right)$ & \textsuperscript{c}$1.5\left(0.03\right)$ & \textsuperscript{c}$0.6\left(0.01\right)$ & \textsuperscript{b}$\textbf{0.0}\left(\textbf{0.00}\right)$ & \textsuperscript{a}$0.3\left(0.01\right)$ \\
    \textbf{CAP} & {\textsuperscript{$\diamond$}ModusGraph} & \textsuperscript{cd}$0.4\left(0.09\right)$ & \textsuperscript{c}$0.3\left(0.09\right)$ & \textsuperscript{c}$9.2\left(2.61\right)$ & \textsuperscript{b}$14.0\left(4.60\right)$ & \textsuperscript{b}$4.0\left(0.70\right)$ & \textsuperscript{a}$\textbf{1.2}\left(\textbf{0.00}\right)$ & \textsuperscript{a}$\textbf{0.8}\left(\textbf{0.00}\right)$ & \textsuperscript{b}$\textbf{0.0}\left(\textbf{0.00}\right)$ & \textsuperscript{a}$0.6\left(0.00\right)$ \\
        & {\textsuperscript{$\diamond$}\textbf{MorphiNet}} & \textsuperscript{a}$\textbf{0.7}\left(\textbf{0.08}\right)$ & \textsuperscript{a}$\textbf{0.6}\left(\textbf{0.09}\right)$ & \textsuperscript{b}$\textbf{6.4}\left(\textbf{3.96}\right)$ & \textsuperscript{a}$\textbf{7.1}\left(\textbf{3.79}\right)$ & \textsuperscript{c}$4.7\left(1.92\right)$ & \textsuperscript{e}$2.4\left(0.38\right)$ & \textsuperscript{e}$0.4\left(0.07\right)$ & \textsuperscript{c}$0.1\left(0.02\right)$ & \textsuperscript{a}$2.7\left(0.39\right)$ \\
        & {\textsuperscript{$\dagger$}NDF} & \textsuperscript{b}\textit{0.6(0.11)} & \textsuperscript{b}\textit{0.5(0.13)} & \textsuperscript{a}\textit{3.4(2.50)} & \textsuperscript{a}\textit{7.5(8.56)} & \textsuperscript{a}\textit{2.1(0.29)} & \textsuperscript{d}\textit{1.7(0.02)} & \textsuperscript{d}\textit{0.6(0.01)} & \textsuperscript{a}\textit{0.0(0.00)} & \textsuperscript{b}\textit{138.7(19.59)} \\

    \hdashline
        \addlinespace[4pt]
        & {\textsuperscript{$\diamond$}HeartDeformNet$\star$} & \textsuperscript{b}$0.4\left(0.07\right)$ & \textbackslash & \textsuperscript{b}$14.1\left(4.58\right)$ & \textbackslash & \textsuperscript{b}$7.2\left(2.57\right)$ & \textsuperscript{b}$\textbf{1.9}\left(\textbf{0.26}\right)$ & \textsuperscript{b}$\textbf{0.5}\left(\textbf{0.06}\right)$ & \textsuperscript{b}$\textbf{0.0}\left(\textbf{0.00}\right)$ & \textsuperscript{a}$\textbf{0.5}\left(\textbf{0.07}\right)$ \\
    \textbf{CAP (LEFT)} & {\textsuperscript{$\diamond$}\textbf{MorphiNet}} & \textsuperscript{a}$\textbf{0.8}\left(\textbf{0.09}\right)$ & \textbackslash & \textsuperscript{a}$\textbf{2.0}\left(\textbf{0.94}\right)$ & \textbackslash & \textsuperscript{a}$\textbf{3.4}\left(\textbf{1.81}\right)$ & \textsuperscript{c}$2.2\left(0.45\right)$ & \textsuperscript{c}$0.4\left(0.08\right)$ & \textsuperscript{c}$0.1\left(0.02\right)$ & \textsuperscript{b}$2.6\left(0.42\right)$ \\
        & {\textsuperscript{$\dagger$}NDF} & \textsuperscript{a}\textit{0.7(0.20)} & \textbackslash & \textsuperscript{a}\textit{2.4(2.24)} & \textbackslash & \textsuperscript{a}\textit{3.1(0.67)} & \textsuperscript{a}\textit{1.6(0.21)} & \textsuperscript{a}\textit{0.6(0.08)} & \textsuperscript{a}\textit{0.0(0.00)} & \textsuperscript{c}\textit{20.2(3.07)} \\
    \hline
        \addlinespace[4pt]
        & {\textsuperscript{$\diamond$}Voxel2Mesh} & \textsuperscript{d}$0.3\left(0.11\right)$ & \textbackslash & \textsuperscript{d}$18.5\left(7.63\right)$ & \textbackslash & \textsuperscript{c}$8.6\left(3.52\right)$ & \textsuperscript{b}$1.4\left(0.04\right)$ & \textsuperscript{b}$0.7\left(0.02\right)$ & \textsuperscript{d}$1.0\left(0.01\right)$ & \textsuperscript{a}$\textbf{0.3}\left(\textbf{0.01}\right)$ \\
        & {\textsuperscript{$\diamond$}CortFlow++} & \textsuperscript{c}$0.4\left(0.08\right)$ & \textbackslash & \textsuperscript{b}$7.2\left(2.28\right)$ & \textbackslash & \textsuperscript{a}$\textbf{3.1}\left(\textbf{0.51}\right)$ & \textsuperscript{bc}$1.5\left(0.03\right)$ & \textsuperscript{c}$0.6\left(0.01\right)$ & \textsuperscript{b}$\textbf{0.0}\left(\textbf{0.00}\right)$ & \textsuperscript{a}$\textbf{0.3}\left(\textbf{0.01}\right)$ \\
    \textbf{ACDC} & {\textsuperscript{$\diamond$}ModusGraph} & \textsuperscript{d}$0.4\left(0.10\right)$ & \textbackslash & \textsuperscript{c}$11.5\left(2.54\right)$ & \textbackslash & \textsuperscript{b}$4.3\left(0.68\right)$ & \textsuperscript{a}$\textbf{1.2}\left(\textbf{0.00}\right)$ & \textsuperscript{a}$\textbf{0.8}\left(\textbf{0.00}\right)$ & \textsuperscript{b}$\textbf{0.0}\left(\textbf{0.00}\right)$ & \textsuperscript{a}$1.4\left(0.02\right)$ \\
        & {\textsuperscript{$\diamond$}\textbf{MorphiNet}} & \textsuperscript{b}$\textbf{0.7}\left(\textbf{0.07}\right)$ & \textbackslash & \textsuperscript{a}$\textbf{2.8}\left(\textbf{1.62}\right)$ & \textbackslash & \textsuperscript{ab}$3.3\left(1.78\right)$ & \textsuperscript{d}$2.4\left(0.57\right)$ & \textsuperscript{e}$0.4\left(0.08\right)$ & \textsuperscript{c}$0.1\left(0.03\right)$ & \textsuperscript{a}$2.3\left(0.26\right)$ \\
        & {\textsuperscript{$\dagger$}NDF} & \textsuperscript{a}\textit{0.7(0.07)} & \textbackslash & \textsuperscript{a}\textit{1.0(0.85)} & \textbackslash & \textsuperscript{a}\textit{2.3(0.59)} & \textsuperscript{c}\textit{1.6(0.02)} & \textsuperscript{d}\textit{0.6(0.01)} & \textsuperscript{a}\textit{0.0(0.00)} & \textsuperscript{b}\textit{133.3(16.64)} \\

    \hdashline
        \addlinespace[4pt]
        & {\textsuperscript{$\diamond$}HeartDeformNet$\star$} & \textsuperscript{b}$0.4\left(0.08\right)$ & \textbackslash & \textsuperscript{b}$16.3\left(3.28\right)$ & \textbackslash & \textsuperscript{b}$7.8\left(2.75\right)$ & \textsuperscript{b}$\textbf{1.9}\left(\textbf{0.30}\right)$ & \textsuperscript{b}$\textbf{0.5}\left(\textbf{0.07}\right)$ & \textsuperscript{b}$\textbf{0.0}\left(\textbf{0.01}\right)$ & \textsuperscript{a}$\textbf{0.6}\left(\textbf{0.09}\right)$ \\
    \textbf{ACDC (LEFT)} & {\textsuperscript{$\diamond$}\textbf{MorphiNet}} & \textsuperscript{a}$\textbf{0.7}\left(\textbf{0.09}\right)$ & \textbackslash & \textsuperscript{a}$\textbf{2.7}\left(\textbf{0.97}\right)$ & \textbackslash & \textsuperscript{a}$\textbf{3.5}\left(\textbf{2.46}\right)$ & \textsuperscript{b}$2.1\left(0.70\right)$ & \textsuperscript{b}$0.4\left(0.11\right)$ & \textsuperscript{c}$0.1\left(0.04\right)$ & \textsuperscript{b}$2.4\left(0.39\right)$ \\
        & {\textsuperscript{$\dagger$}NDF} & \textsuperscript{a}\textit{0.8(0.12)} & \textbackslash & \textsuperscript{a}\textit{1.4(1.53)} & \textbackslash & \textsuperscript{a}\textit{2.8(0.73)} & \textsuperscript{a}\textit{1.6(0.24)} & \textsuperscript{a}\textit{0.6(0.09)} & \textsuperscript{a}\textit{0.0(0.00)} & \textsuperscript{c}\textit{27.2(4.73)} \\
    \hline
    \end{tabularx}
    \vspace{2mm}
    \begin{minipage}{\textwidth}
    \footnotesize
    \textit{Notes:} Dice and Hd are calculated based on pixels for SAX and LAX views. ASD is in mm. AspR, JacR, and NmF are dimensionless. InFt is in seconds. CAP (LEFT) and ACDC (LEFT) indicate evaluation done solely with left-ventricular myocardium. For ACDC, LAX scores are unavailable due to a lack of LAX data and denoted by `\textbackslash'. For CAP (LEFT), LAX scores are unavailable due to missing left two-chamber data and denoted by `\textbackslash'. Numbers represent average values and (standard deviation).
    \end{minipage}
\end{table*}

\subsection{Motion Tracking Evaluation}
To assess MorphiNet's capability in capturing cardiac motion, we conducted a comparative study tracking the deformation from end-diastole (ED) to end-systole (ES) against BiVModel \cite{maugerIterativeDiffeomorphicAlgorithm2018}, an established method for biventricular model template registration. Crucially, BiVModel reconstructs meshes directly from ground-truth segmentations rather than raw images, thereby serving as a \textit{high target benchmark} for geometric accuracy and functional estimation rather than a baseline for automated inference. 

To quantify the physiological realism of the reconstructed motion, we employ two smoothness metrics. \textit{Max Curvature} (Curv) measures the geometric complexity of vertex trajectories, while \textit{Max Jerk} quantifies the rate of change of acceleration over $T$ time frames:
\begin{equation}
\begin{aligned}
    \text{Curv} &= \max_{v,t} \frac{\|\bm{v}_{t} \times \bm{a}_{t}\|}{\|\bm{v}_{t}\|^3} \\
    \text{Jerk} &= \max_{v \in \mathcal{V}} \frac{1}{T-3} \sum_{t=0}^{T-3} \left\| \frac{\partial^3 \bm{p}_{v,t}}{\partial t^3} \right\|_2
\end{aligned}
\end{equation}
where $\bm{v}_{t}$, $\bm{a}_{t}$, and $\bm{p}_{v,t}$ denote the velocity, acceleration, and position vectors of vertex $v$ at time $t$.

\begin{figure*}[!t]
  \centering
  % --- Left (wide) subfigure ---
  \begin{subfigure}[t]{0.74\textwidth}
    \includegraphics[width=\linewidth,keepaspectratio]{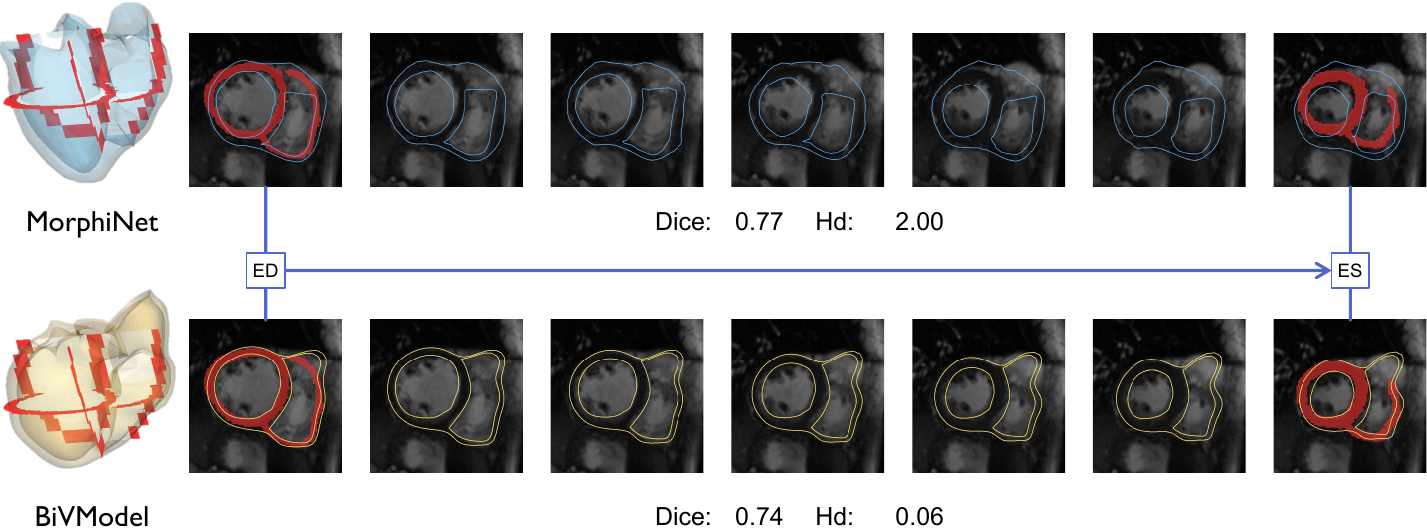}
    \caption{}
    \label{fig:4d_left}
  \end{subfigure}
  \hfill
  % --- Right (narrow) subfigure ---
  \begin{subfigure}[t]{0.175\textwidth}
    \includegraphics[width=\linewidth,keepaspectratio]{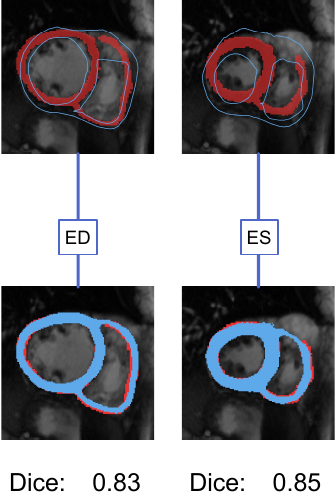}
    \caption{}
    \label{fig:4d_right}
  \end{subfigure}

  \caption{\textbf{Time-resolved reconstruction from end-diastole (ED) to end-systole (ES).} (a) Mid-slice short-axis (SAX) overlays of reconstructed meshes against CMR images and ground truth segmentations (\textcolor{red}{red}). The first row displays MorphiNet reconstructions (\textcolor{blue}{blue}), while the second row displays BiVModel reconstructions (\textcolor{yellow}{yellow}). Dice scores and Hausdorff Distances (Hd), calculated as described in Section~\ref{sec:surface_reconstruction_performance}, are reported for this representative case. (b) Assessment of segmentation accuracy: the first row overlays MorphiNet's reconstructed model against the ground truth; the second row contrasts dynUNet's segmentation prediction (\textcolor{blue}{blue}) with the ground truth. Note that BiVModel utilizes ground truth segmentation for reconstruction, serving here as a reference benchmark.}
  \label{fig:4d_morphing}
\end{figure*}

Fig.~\ref{fig:4d_morphing} and Table~\ref{table:morph_quality} present the tracking results. We use linear interpolation between frames because BiVModel does not predict a mesh at every frame, which results in zero curvature and constant acceleration (zero jerk). Thus, we show only the linear change for BiVModel in Fig. \ref{fig:4d_morphing}, and omit its curvature and acceleration scores from Table \ref{table:morph_quality}. As expected, the benchmark BiVModel achieves superior geometric accuracy (Dice 0.74, ASD 1.84mm) due to its reliance on segmentation ground truth. MorphiNet exhibits a geometric gap (Dice 0.67, ASD 3.18mm), primarily attributed to the low resolution ($16 \times 16 \times 16$) of the gradient field, which limits the deformation's ability to approximate a fully diffeomorphic mapping. The large variance in Curvature values indicates that MorphiNet lacks temporal constraints for robust, smooth motion tracking. Despite this, MorphiNet demonstrates a faithful reflection of cardiac function, achieving mean ejection fractions (LV: 37.79\%, RV: 33.64\%) statistically comparable to the benchmark (LV: 47.01\%, RV: 38.96\%), enabling effective clinical stratification that correctly identifies myocardial dysfunction in tetralogy of Fallot patients. 

\begin{table}[ht]
    \centering
    \caption{\textbf{Motion tracking evaluation on CAP dataset.} MorphiNet yields comparable mean functional metrics to the benchmark despite lower geometric precision. All comparable metrics show statistically significant differences ($p<0.001$), calculated via paired t-test on 13 subjects.  \textsuperscript{$\dagger$}BiVModel utilizes ground truth segmentation for reconstruction and serves as a high target benchmark.}
    \label{table:morph_quality}
    \begin{tabularx}{\columnwidth}{l|CC}
        \hline
        \textbf{Metric} & \textsuperscript{$\dagger$}\textbf{BiVModel} & \textbf{MorphiNet} \\
        \hline

        \addlinespace[4pt]

        Dice$\uparrow$ & $0.74\left(0.09\right)$ & $0.67\left(0.08\right)$ \\
        Hd$\downarrow$ & $0.07\left(0.02\right)$ & $2.58\left(0.96\right)$ \\
        ASD$\downarrow$ & $1.84\left(0.23\right)$ & $3.18\left(1.12\right)$ \\
        AspR$\downarrow$ & $1.39\left(0.02\right)$ & $2.08\left(0.17\right)$ \\
        JacR$\uparrow$ & $0.73\left(0.02\right)$ & $0.45\left(0.04\right)$ \\
        NmF$\downarrow$ & $0.01\left(0.00\right)$ & $0.07\left(0.01\right)$ \\
        Max Curv$\downarrow$ & \textbackslash & $403.45\left(749.99\right)$ \\
        Max Jerk$\downarrow$ & \textbackslash & $0.05\left(0.01\right)$ \\ \hline

        \addlinespace[4pt]

        LV EF (\%) & $47.01\left(5.96\right)$ & $37.79\left(10.03\right)$ \\
        RV EF (\%) & $38.96\left(8.76\right)$ & $33.64\left(12.46\right)$ \\
        \hline
    \end{tabularx}
    \vspace{2mm}
    \begin{minipage}{\columnwidth}
    \footnotesize
    \textit{Notes:} Numbers represent average values and (standard deviation). All values are dimensionless unless specified.
    \end{minipage}
\end{table}

\subsection{Ablation Studies}
\subsubsection{Segmentation Analysis} We evaluated the accuracy of the \textbf{segmentation completion}, defined as the combination of the \textbf{partial segmentation} provided by dynUNet and the \textbf{complementary segmentation} predicted by ResNet. This analysis spanned both CT (SCOT-HEART) and CMR (CAP) data to assess the fidelity of the restored anatomical structures.

For CT data, we employed a masking strategy in which slices proximal to the basal and apical planes were intentionally removed to mimic the commonly missing segmentations in typical CMR acquisitions. The ResNet was then tasked with predicting a complementary segmentation for these missing regions. The quantitative results, reported in the 'CT (after)' column of Table \ref{table:ct_mr_segmentation}, demonstrate the network's capacity to restore the anatomy from partial data, effectively ``guessing'' the missing structure to match the complete manual segmentation.

For CMR data, we evaluated performance at both ED and ES phases. While standard metrics (Dice and Hd) remained consistent, specific challenges emerged in the volumetric analysis. We observed a systematic volume overestimation in the ES phase after ResNet restoration, as evidenced by the significant volume difference (Vol Diff \%) in Table \ref{table:ct_mr_segmentation}. This abnormality is attributed to the learning mechanism; the ResNet learns primarily from CT data, which is typically acquired in a phase resembling diastole. Consequently, the complementary segmentation tends to impose a diastolic-like morphology, resulting in ``overestimated'' wall thickness and cavity volume when applied to ES geometry. Furthermore, the myocardium proved more challenging to complete than the LV or RV blood pools, likely due to the thinness of the RV wall and lower tissue contrast.

\begin{figure}[!t]
    \includegraphics[width=\columnwidth]{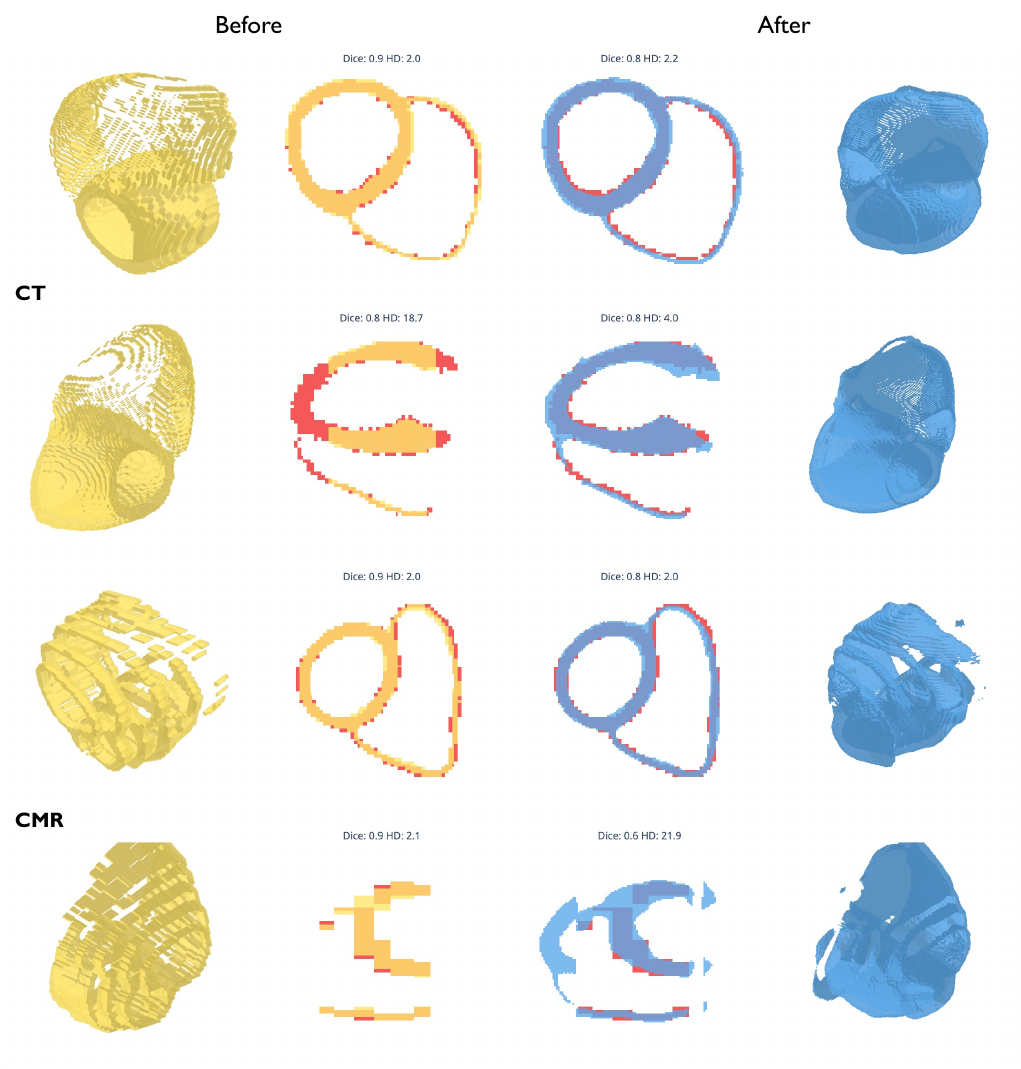}
    \caption{\textbf{Visualization of segmentation impact driven by ResNet.} The first two rows depict a representative CT case (SCOT-HEART), while the last two rows depict a representative CMR case (CAP). Across the columns, the first two columns display the \textit{partial segmentation} (\textcolor{yellow}{yellow}) produced by dynUNet (``before''), while the last two columns display the \textit{segmentation completion} (\textcolor{blue}{blue}) (``after''). The ground truth is overlaid in \textcolor{red}{red} for comparison.}
    \label{fig:effect_resnet}
\end{figure}

To visualize the impact of ResNet restoration, Fig. \ref{fig:effect_resnet} contrasts the partial segmentation with the segmentation completion. It is crucial to highlight the role of ResNet in achieving superior anatomical fidelity. While the CMR dynUNet processes full stacks without cropping, its partial segmentation undergoes nearest-neighbour interpolation to match the target resolution, often resulting in jagged boundaries. In contrast, the ResNet utilizes the \textbf{feature map} directly from the dynUNet to ``hallucinate'' the missing anatomical structure. By comparing the results in Fig. \ref{fig:effect_resnet}, it is evident that the segmentation completion exhibits significantly smoother boundaries than the nearest-neighbour interpolated partial segmentation, despite the latter achieving high accuracy scores in the 'before' columns of Table \ref{table:ct_mr_segmentation}. This confirms ResNet's contribution to providing a complete, anatomically coherent structure.

Finally, it is worth noting that MorphiNet acts as a modular framework. While dynUNet and ResNet were selected here for computational feasibility, they serve as interchangeable modules. Future implementations may substitute these with alternative networks to address specific challenges, such as systolic volume estimation, without altering the core pipeline.
\begin{table}[!t]
    \centering
    \small
    \setlength{\tabcolsep}{3pt}
    \caption{\textbf{Quantitative evaluation of dynUNet segmentation accuracy}, comparing \textit{partial segmentation} (before) and \textit{segmentation completion} (after). The segmentation completion maintains consistent geometric accuracy (Dice/Hd) while significantly restoring missing anatomical volume, despite a systematic overestimation in the ES phase due to diastolic training bias. Statistical significance is derived from paired t-tests.}
    \label{table:ct_mr_segmentation}
    \begin{tabularx}{\columnwidth}{L L|C|C C C|C C C}
        \hline
        \textbf{Metric} & & \textbf{CT} & \multicolumn{3}{c}{\textbf{CMR-ED}} & \multicolumn{3}{c}{\textbf{CMR-ES}} \\
        &  & after & before & after & $p$ & before & after & $p$ \\
        \hline
        \addlinespace[4pt]  % Adds 4pt of vertical space
        & LV & 0.94 & 0.93 & 0.89 & {\tiny $<$0.001} & 0.90 & 0.85 & {\tiny $<$0.001} \\
        \textbf{Dice} & MYO & 0.83 & 0.79 & 0.72 & {\tiny $<$0.001} & 0.80 & 0.75 & {\tiny $<$0.001} \\
        & RV & 0.93 & 0.92 & 0.90 & {\tiny $<$0.001} & 0.89 & 0.87 & {\tiny $<$0.05} \\ 
        \hline
        \addlinespace[4pt]  % Adds 4pt of vertical space
        & LV & 3.01 & 4.67 & 6.23 & {\tiny $\geq$0.05} & 6.07 & 11.68 & {\tiny $<$0.001} \\
        \textbf{Hd} & MYO & 3.05 & 3.20 & 3.16 & {\tiny $\geq$0.05} & 4.01 & 3.67 & {\tiny $\geq$0.05} \\
        & RV & 3.16 & 4.70 & 3.97 & {\tiny $<$0.01} & 7.88 & 5.27 & {\tiny $<$0.05} \\ 
        \hline
        \addlinespace[4pt]  % Adds 4pt of vertical space
        \textbf{Vol} & LV & 18.95 & \textbackslash & 8.40 & \textbackslash & \textbackslash & 8.80 & \textbackslash \\
        \textbf{Diff} & MYO & 25.78 & \textbackslash & 8.27 & \textbackslash & \textbackslash & 4.70 & \textbackslash \\
        \% & RV & 12.98 & \textbackslash & 1.26 & \textbackslash & \textbackslash & 1.19 & \textbackslash \\
        \hline
    \end{tabularx}
    \vspace{2mm}
    \begin{minipage}{\columnwidth}
    \footnotesize
    \textit{Notes:} Numbers represent average values; all values are dimensionless.
    \end{minipage}
\end{table}

\subsubsection{Deformation Strategy Comparison}
\label{sec:mesh_quality_worsening}
We conducted an ablation study to isolate the contributions of the gradient-field deformation and the adaptive subdivision. Using identical segmentation predictions from dynUNet, we compared five strategies: \circnum{1} \textbf{Base Model + Loop Subdivision} (applying Loop subdivision to the undeformed template); \circnum{3} \textbf{Gradient Deformation} (adjusting the template via the gradient field without subdivision); \circnum{2} \textbf{Gradient Deformation + Loop Subdivision} (applying standard Loop subdivision after deformation); and our proposed adaptive approaches, \circnum{4} \textbf{Gradient Deformation + Single GSN} and \circnum{5} \textbf{Gradient Deformation + Double GSN}.

Table \ref{table:deformation-strategies} quantifies the benefits of the proposed components. The gradient deformation is fundamental, raising the Dice score from 0.56 (\circnum{1}) to 0.78 (\circnum{3}). Crucially, the adaptive GSN proves superior to standard subdivision; compared to using Loop subdivision (\circnum{2}), our double-layer GSN (\circnum{5}) achieves higher geometric fidelity (Dice 0.83 vs. 0.76) and significantly lower surface error (Hd 1.47 vs. 2.17).

\begin{figure}[!t]
  \centering
  % --- Row 1 ---
  \begin{subfigure}[t]{0.175\textwidth}
    \centering
    \includegraphics[width=\linewidth,keepaspectratio]{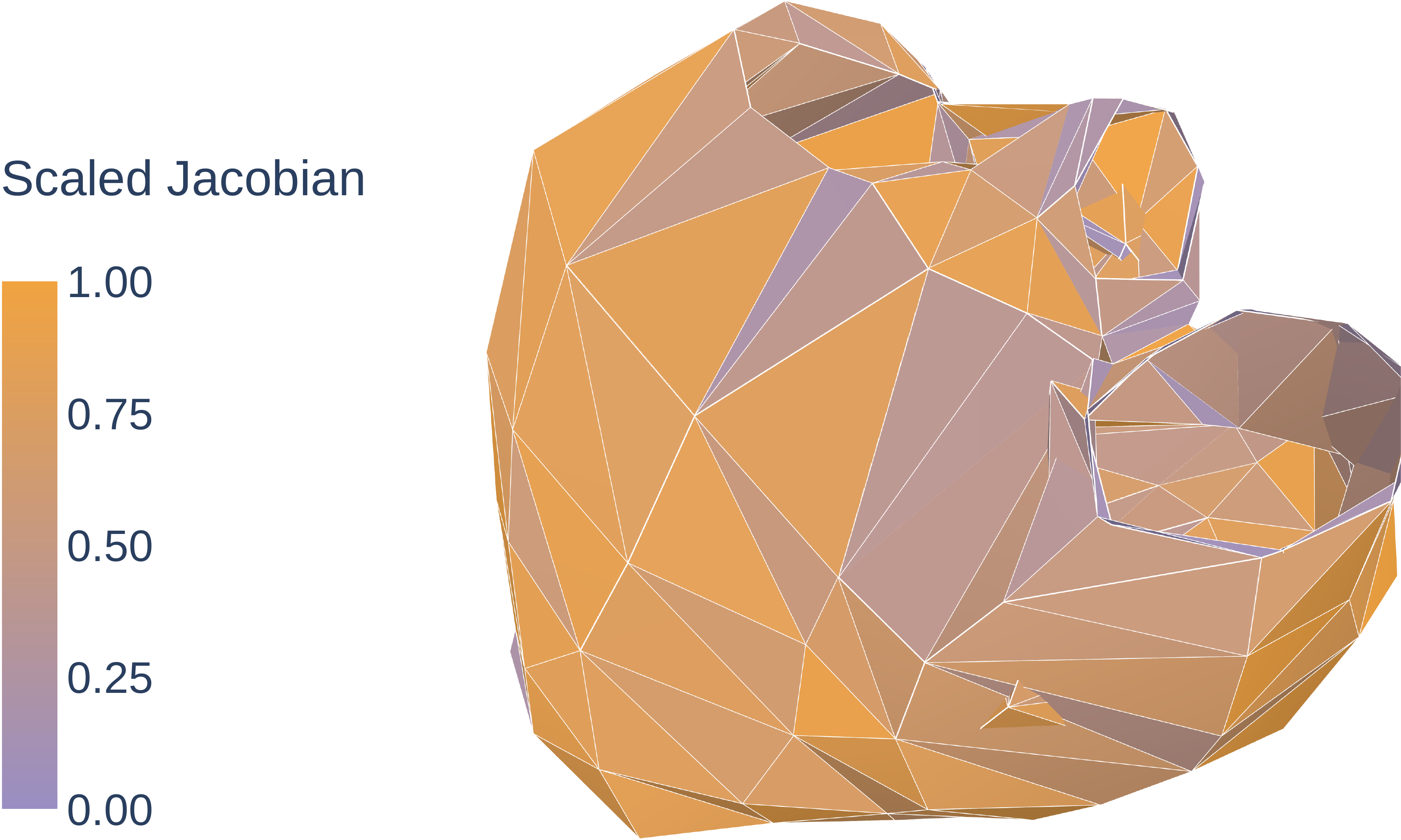}
    \caption{\circnum{3} - JacR}
  \end{subfigure}
  \hfill
  \begin{subfigure}[t]{0.12\textwidth}
    \centering
    \includegraphics[width=\linewidth,keepaspectratio]{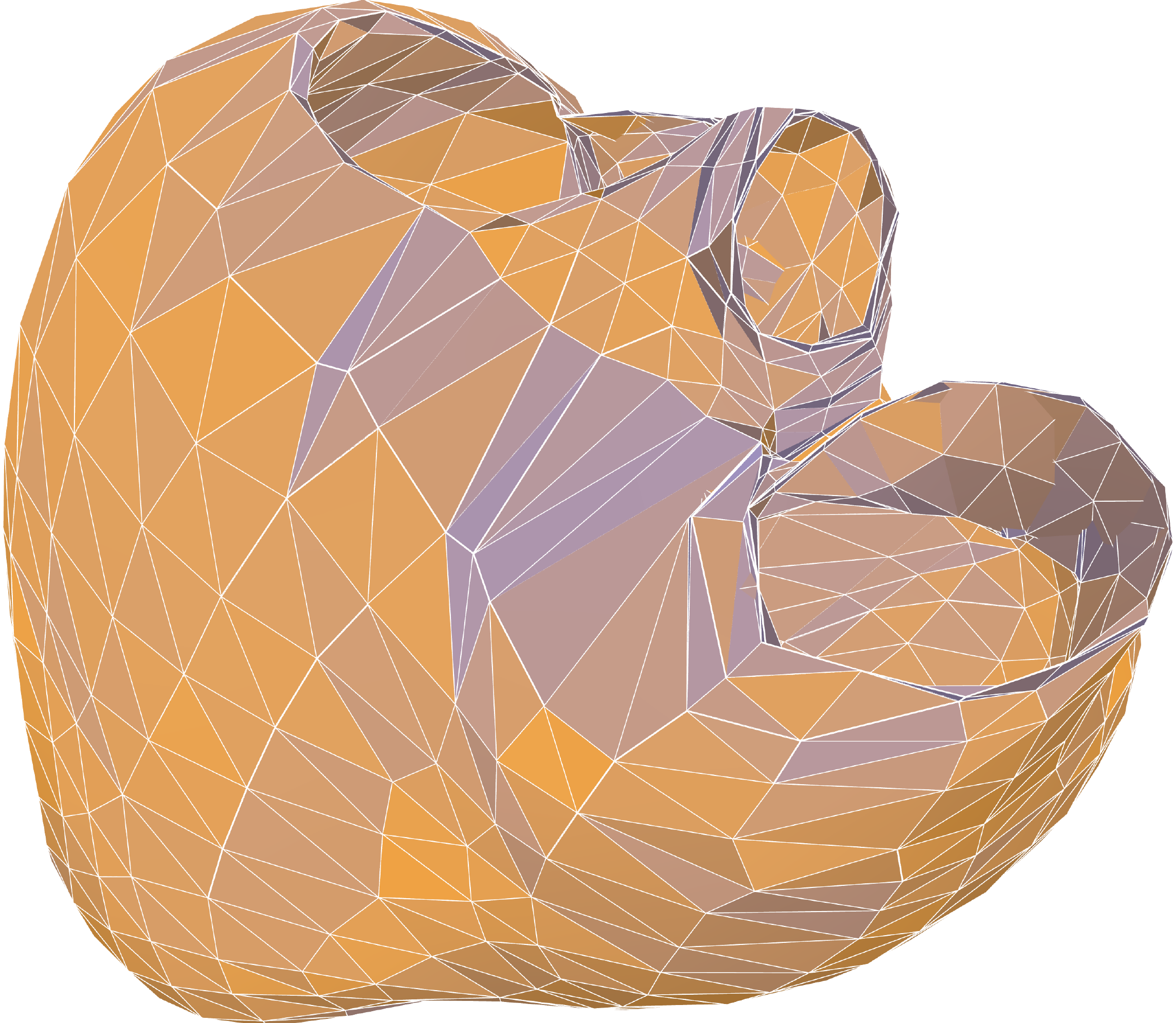}
    \caption{\circnum{4} - JacR}
  \end{subfigure}
  \hfill
  \begin{subfigure}[t]{0.12\textwidth}
    \centering
    \includegraphics[width=\linewidth,keepaspectratio]{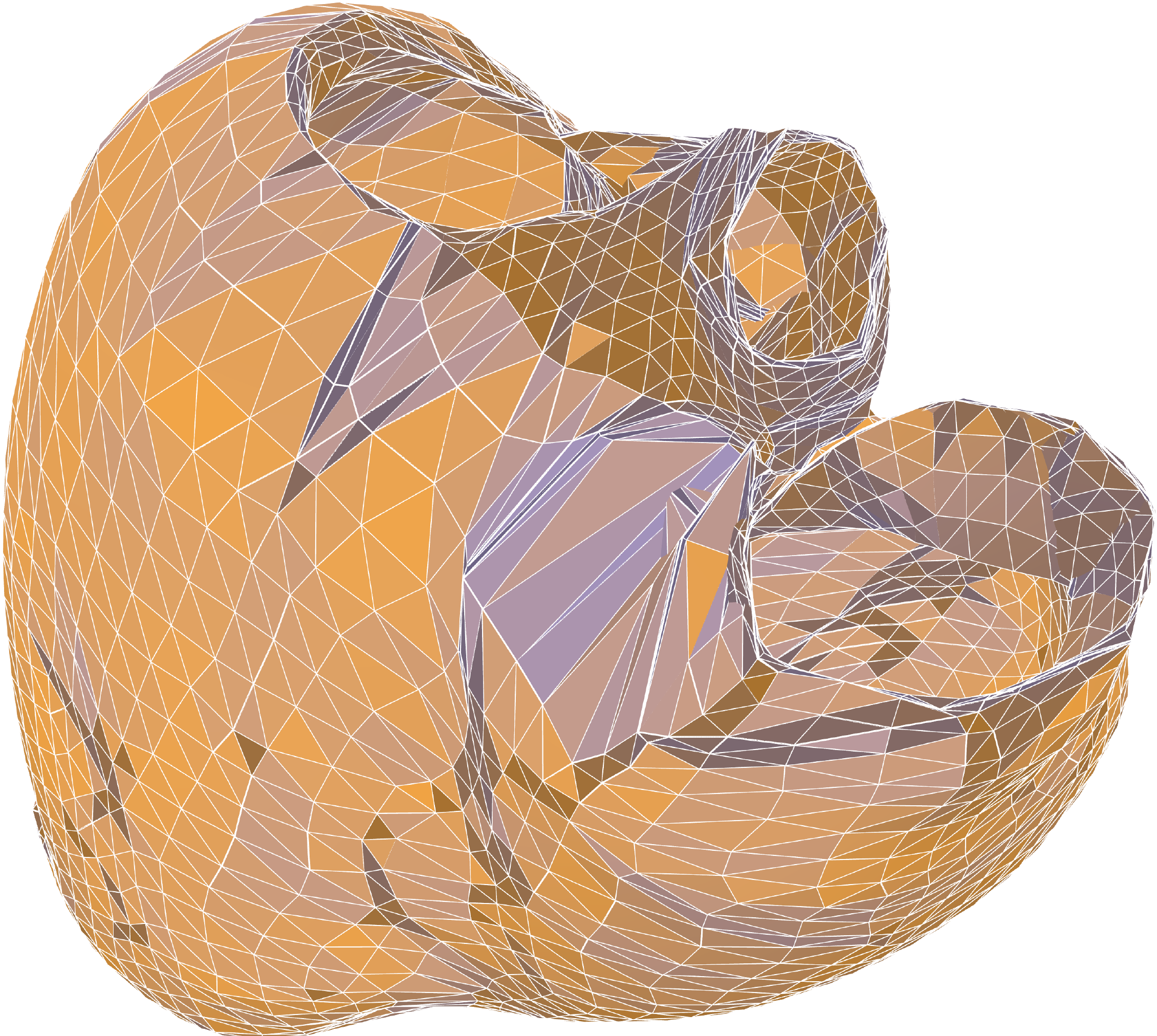}
    \caption{\circnum{5} - JacR}
  \end{subfigure}

  \par\medskip   % <--- start second row

  % --- Row 2 ---
  \begin{subfigure}[t]{0.175\textwidth}
    \centering
    \includegraphics[width=\linewidth,keepaspectratio]{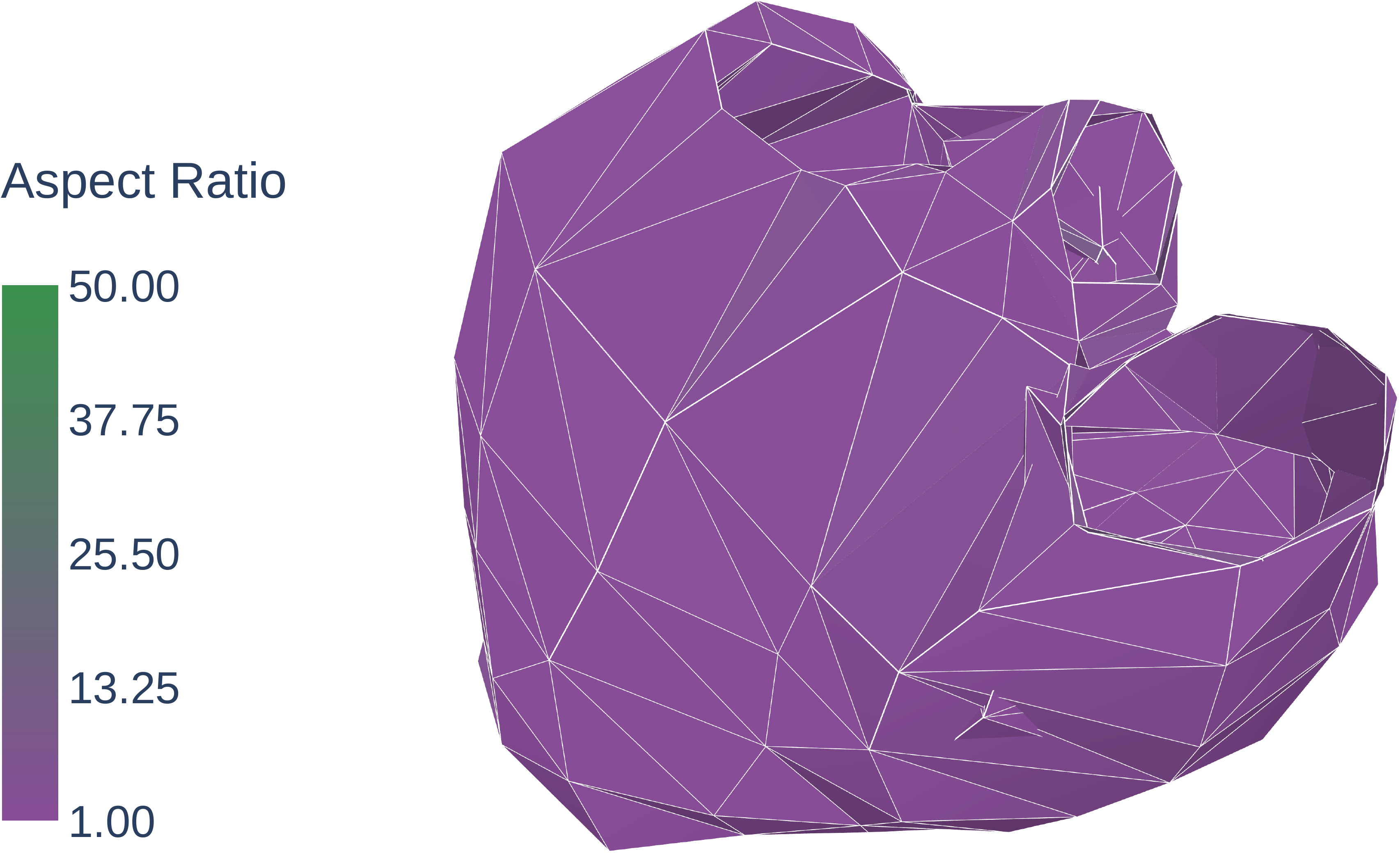}
    \caption{\circnum{3} - AspR}
  \end{subfigure}
  \hfill
  \begin{subfigure}[t]{0.12\textwidth}
    \centering
    \includegraphics[width=\linewidth,keepaspectratio]{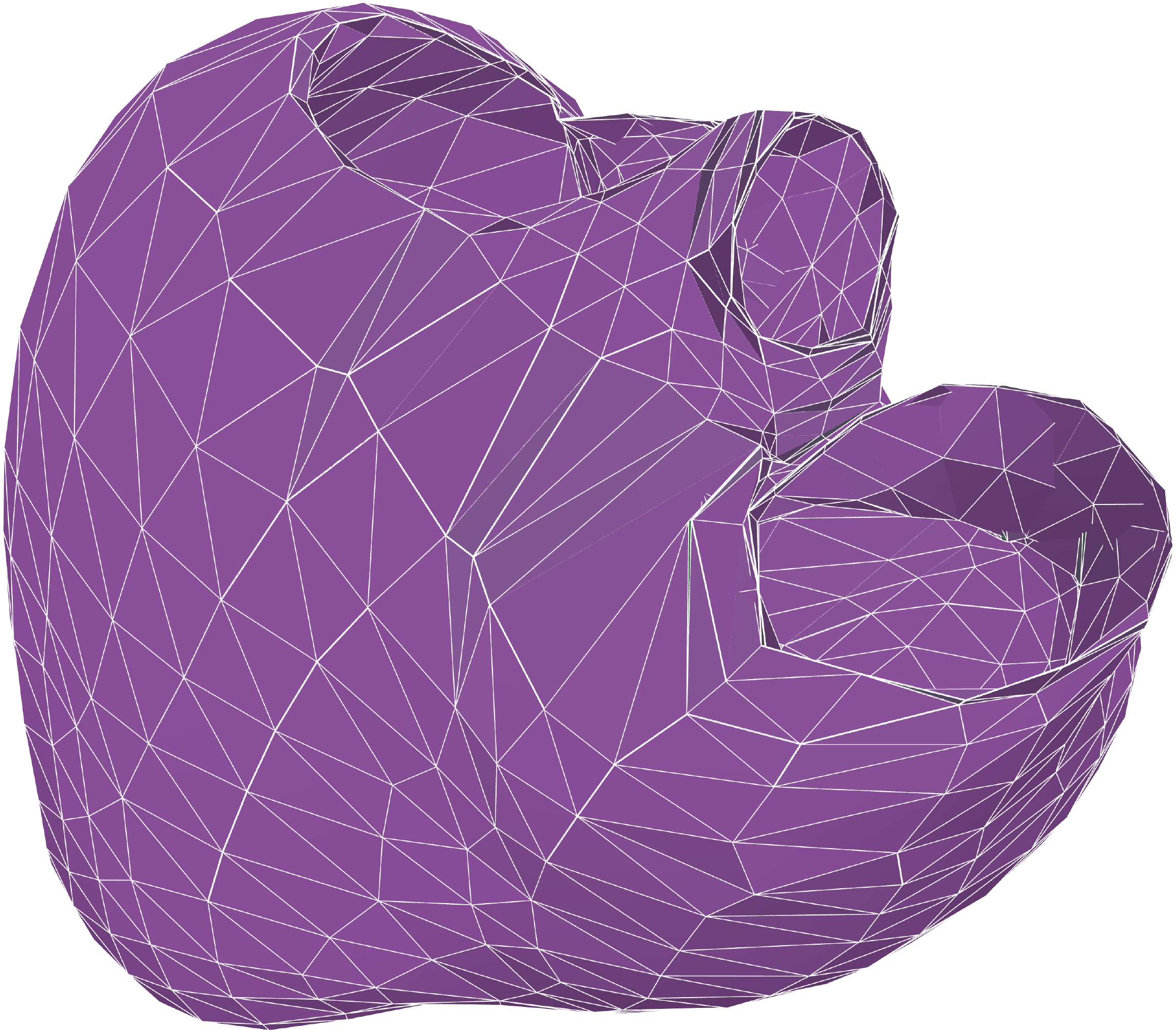}
    \caption{\circnum{4} - AspR}
  \end{subfigure}
  \hfill
  \begin{subfigure}[t]{0.12\textwidth}
    \centering
    \includegraphics[width=\linewidth,keepaspectratio]{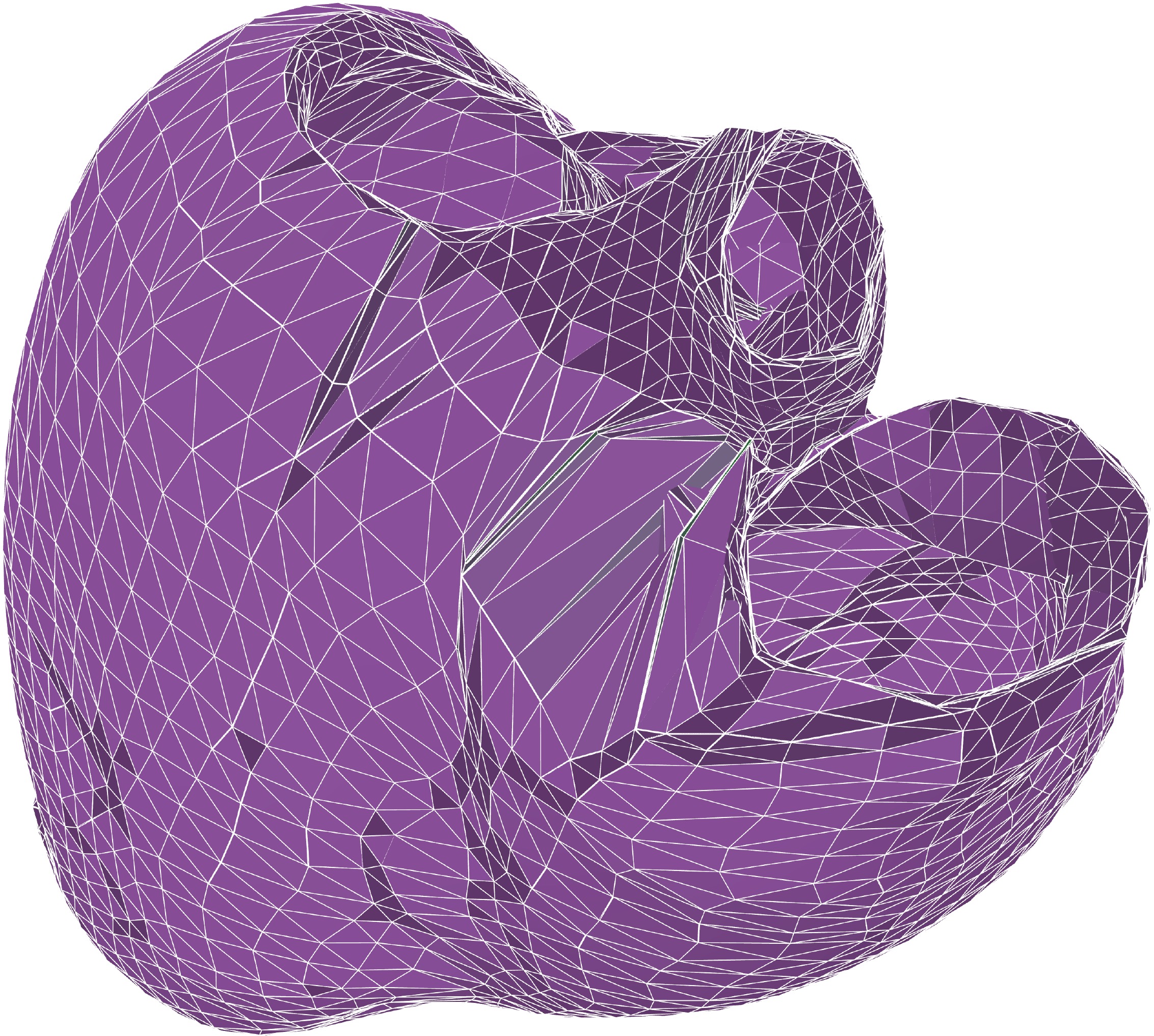}
    \caption{\circnum{5} - AspR}
  \end{subfigure}
  \caption{\textbf{Progressive impact of adaptive subdivision on mesh element quality.} The columns display the mesh state at three stages: (Left) immediately after Gradient Deformation (\circnum{3}); (Middle) after one GSN layer (\circnum{4}); and (Right) after two GSN layers (\circnum{5}). The first row visualizes the Scaled Jacobian Ratio (JacR), and the second row visualizes the Aspect Ratio (AspR). While the face count increases to improve geometric fit, a degradation in element quality is observed, particularly in the thin, triangulated regions surrounding the valves.}
  \label{fig:mesh_label_quality}
\end{figure}

However, this geometric refinement incurs a trade-off in element regularity. As shown in the `$<$0.7' column, the undeformed template starts with 36\% of faces exhibiting JacR below 0.7. The adaptive subdivision process further reduces the mean JacR from 0.62 (\circnum{3}) to 0.59 (\circnum{5}). Fig.~\ref{fig:mesh_label_quality} visualizes this progressive change: while the number of faces increases to capture fine details, element quality metrics (JacR and AspR) show degradation. This worsening is spatially concentrated around thin structures, such as the tricuspid valve. In these regions, the template --- originally defined with quadrilateral faces --- was triangulated to ensure compatibility with deep learning frameworks (\texttt{PyTorch3D} and \texttt{Pytorch-Geometric} used in the study). This triangulation of rectangular geometry creates inherently skewed elements, the irregularity of which is exacerbated during the subdivision process. Future work will aim to develop subdivision schemes that maintain the element regularity inherent to Loop subdivision, while optimizing the trade-off between anatomical fidelity and mesh shrinkage to prevent volume loss.
\begin{table}[!t]
    \centering
    \small
    \setlength{\tabcolsep}{1pt} % Optional: Adjust column separation if needed
    \caption{\textbf{Quantitative comparison of geometric accuracy and mesh quality on CT data (SCOT-HEART)}, validating the double GSN layer design (\circnum{5}) for superior anatomical fidelity despite element regularity trade-offs. Strategies include: \circnum{1} Base Model + Loop Subdivision, \circnum{2} Gradient Deformation + Loop Subdivision, \circnum{3} Gradient Deformation, \circnum{4} Gradient Deformation + Single GSN, and \circnum{5} Gradient Deformation + Double GSN. `$<$0.7' denotes the ratio of faces with JacR $<$ 0.70. Superscript letters denote Compact Letter Display (CLD) groupings: methods sharing a letter are not significantly different ($p \ge 0.05$), with `a' representing the best performance.}
    \label{table:deformation-strategies}
    \begin{tabularx}{\columnwidth}{C|C C C C C C C C}
    \hline
        & \textbf{Dice}$\uparrow$ & \textbf{Hd}$\downarrow$ & \textbf{ASD}$\downarrow$ & \textbf{AspR}$\downarrow$ & \textbf{JacR}$\uparrow$ & \textbf{$<$0.7}$\downarrow$ & \textbf{MnC}$\uparrow$ & \textbf{NmF}$\downarrow$ \\
    \hline
        \addlinespace[4pt]  % Adds 4pt of vertical space
        \circnum{1} & \textsuperscript{d}0.56 & \textsuperscript{d}4.38 & \textsuperscript{b}3.09 & \textsuperscript{a}\textbf{1.35} & \textsuperscript{a}\textbf{0.76} & \textsuperscript{a}\textbf{0.36} & \textsuperscript{b}0.76 & \textsuperscript{a}\textbf{0.00} \\ 
        \circnum{2} & \textsuperscript{c}0.76 & \textsuperscript{c}2.17 & \textsuperscript{a}1.56 & \textsuperscript{b}1.38 & \textsuperscript{b}0.73 & \textsuperscript{b}0.45 & \textsuperscript{a}\textbf{0.78} & \textsuperscript{b}0.01 \\ 
    \hdashline
        \addlinespace[4pt]  % Adds 4pt of vertical space
        \circnum{3} & \textsuperscript{b}0.78 & \textsuperscript{b}1.84 & \textsuperscript{a}1.40 & \textsuperscript{d}1.60 & \textsuperscript{d}0.62 & \textsuperscript{d}0.62 & \textsuperscript{b}0.76 & \textsuperscript{d}0.15 \\
        \circnum{4} & \textsuperscript{a}0.82 & \textsuperscript{ab}1.59 & \textsuperscript{a}1.53 & \textsuperscript{c}1.54 & \textsuperscript{c}0.63 & \textsuperscript{c}0.60 & \textsuperscript{a}\textbf{0.78} & \textsuperscript{c}0.04 \\
        \circnum{5} & \textsuperscript{a}\textbf{0.83} & \textsuperscript{a}\textbf{1.47} & \textsuperscript{a}\textbf{1.35} & \textsuperscript{e}1.64 & \textsuperscript{e}0.59 & \textsuperscript{e}0.65 & \textsuperscript{a}\textbf{0.78} & \textsuperscript{c}0.04 \\
    \hline
    \end{tabularx}
    \vspace{2mm}
    \begin{minipage}{\columnwidth}
    \footnotesize
    \textit{Notes:} Dice and Hd are pixel-based; ASD is in mm; AspR, JacR, MnC, and NmF are dimensionless. Numbers represent average values, with best scores in bold.
    \end{minipage}
\end{table}

\section{Discussion}
MorphiNet demonstrates significant advantages in heart mesh reconstruction, effectively balancing anatomical fidelity with computational efficiency. MorphiNet achieves Dice scores of 0.8 on CT data. With a single RTX 3090 GPU, we trained MorphiNet in 14 hours and obtained a 0.3 higher Dice score and 2.6 lower Hausdorff distance improvement over the previous best template-based method on the CAP dataset. This performance is equivalent to the top-performing neural implicit function, but at a much lower inference cost---2.7 seconds per patient compared with 138.7 seconds by the neural implicit function. Recognizing the domain shift caused by protocols, vendors, and patients differing from cardiac imaging data sources, we employed zero-shot validation on the ACDC. MorphiNet obtains 0.7 Dice score, 2.8 Hausdorff distance and 3.3~mm ASD, which is the same level of performance as in the CAP dataset and is significantly better than the best template-based method. This suggests that MorphiNet performs better than expected, and its learned anatomical structure is general.

While MorphiNet's template-based vertex displacement offers superior topology control, it presents a trade-off against neural implicit functions like NDF. NDF achieves a lower ASD (1.0 mm vs. 1.4 mm for MorphiNet on CT) through pixel-wise optimization. However, this comes at a prohibitive computational cost (133.1 s vs. 2.7 s for MorphiNet) and results in meshes lacking point correspondence. By maintaining a fixed triangular topology and consistent vertex count (expanding from 388 to 6,238 vertices), MorphiNet ensures the temporal consistency required for dynamic strain analysis and motion tracking, a capability often absent in neural implicit functions.

Crucially, MorphiNet demonstrates robust functional assessment capabilities, overcoming the challenges of domain shift. In our motion tracking evaluation, the method yields ejection fraction predictions (LV: 37.8\%, RV: 33.6\%) that are statistically comparable to the ground-truth benchmark BiVModel (LV: 47.0\%, RV: 39.0\%). This contradicts the potential concern that the ResNet's training on static CT data might impair dynamic function estimation. Although our ablation study (Table \ref{table:ct_mr_segmentation}) indicated that the ResNet's complementary segmentation tends to overestimate end-systolic volumes due to a bias toward diastolic morphologies learned from CT, the subsequent gradient deformation and GSN refinement steps appear to mitigate this effect in the final mesh. Consequently, MorphiNet successfully captures the non-linear dynamics of the heart—evidenced by non-zero Jerk and Curvature metrics—allowing for the correct identification of myocardial dysfunction in clinical cohorts such as tetralogy of Fallot.

The ablation studies further reveal a complex relationship between anatomical fidelity and mesh element quality. The introduction of the Adaptive Subdivision Process is fundamental to MorphiNet's accuracy; the double-layer GSN improves the Dice score from 0.78 to 0.83 on CT data compared to deformation alone. However, this geometric refinement induces a trade-off in element regularity. As the mesh adapts to fine anatomical details, we observe a degradation in quality metrics, with the JacR decreasing from 0.62 to 0.59. Surface discrepancies are also noted near the basal plane. These discrepancies occur because the input segmentation lacks specific valvular landmarks. Consequently, the gradient field deformation in these open-boundary regions is driven by general surface proximity rather than precise anatomical anchor points, leading to higher surface errors compared to the pixel-perfect fit of neural implicit functions.

While MorphiNet advances template-based methods, several limitations delineate the path for future improvement. First, the ResNet-based segmentation completion exhibits a domain bias; trained primarily on CT data (typically acquired in diastole), it tends to impose diastolic morphologies on systolic CMR frames, leading to the volume overestimation observed in our segmentation analysis. Future work could incorporate cycle-consistent training or diverse cardiac phase data to mitigate this bias. Second, the adaptive GSN refinement introduces a trade-off between geometric fidelity and element quality, particularly in thin, triangulated regions like the valve plane where aspect ratios worsen. Developing isotropic remeshing techniques or incorporating element-quality regularization terms into the GSN loss function will be essential to prevent mesh quality degradation. Finally, the absence of specific valve landmarks in the input segmentation limits the gradient field's ability to precisely delineate valvular boundaries. Integrating multi-chamber segmentation that includes atrial and arterial roots could provide the necessary boundary conditions for more accurate valve localization.

\section{Conclusion}
In this paper, we address the challenges in heart model reconstruction and modeling due to the anisotropic characteristics of CMR imaging while acknowledging the complementary anatomical insights offered by CT. We propose MorphiNet, a GSN that predicts a continuous gradient field and adaptively adjusts and refines the mesh in a patient-specific manner. Compared with state-of-the-art methods, empirical validation on CMR and CT datasets substantiates MorphiNet's performance in capturing complex heart anatomical structures, demonstrating its potential to generate high-fidelity heart mesh models.

\end{document}